\documentclass[journal,twocolumn]{IEEEtran}
\usepackage{amsfonts}
\usepackage{times}
\usepackage{graphicx}
\usepackage{latexsym}
\usepackage{dsfont}
\usepackage{amssymb}
\usepackage{amsmath}
\usepackage{cite}
\usepackage{verbatim}
\usepackage{subfigure}

\newcommand{\figref}[1]{{Fig.}~\ref{#1}}

\def\bb0{{\mathbb{0}}}

\def\ba{{\mathbf{a}}}
\def\bb{{\mathbf{b}}}

\def\bff{{\mathbf{f}}}

\def\bh{{\mathbf{h}}}

\def\bn{{\mathbf{n}}}

\def\bp{{\mathbf{p}}}

\def\bu{{\mathbf{u}}}
\def\bv{{\mathbf{v}}}

\def\bx{{\mathbf{x}}}
\def\by{{\mathbf{y}}}

\def\b0{{\mathbf{0}}}

\def\bA{{\mathbf{A}}}
\def\bB{{\mathbf{B}}}

\def\bD{{\mathbf{D}}}
\def\bE{{\mathbf{E}}}
\def\bF{{\mathbf{F}}}
\def\bG{{\mathbf{G}}}
\def\bH{{\mathbf{H}}}
\def\bI{{\mathbf{I}}}

\def\bN{{\mathbf{N}}}

\def\bP{{\mathbf{P}}}
\def\bQ{{\mathbf{Q}}}

\def\bU{{\mathbf{U}}}

\def\bX{{\mathbf{X}}}
\def\bY{{\mathbf{Y}}}
\def\bZ{{\mathbf{Z}}}

\def\bbE{{\mathbb{E}}}

\def\cA{\mathcal{A}}

\def\sf0{{\mathsf{0}}}

\usepackage{epstopdf}
\usepackage{enumerate}
\usepackage{algorithm}
\usepackage{amsmath}
\usepackage{float}
\usepackage{color}
\usepackage{makeidx}
\usepackage{bm}
\usepackage[hidelinks=true]{hyperref}
\usepackage{cleveref}
\usepackage{url}
\usepackage{steinmetz}
\usepackage{varwidth}% http://ctan.org/pkg/varwidth
\usepackage{soul}
\usepackage{lipsum}
\usepackage{cuted} 
\usepackage{balance}
\newcommand{\sref}[1]{{Section}~\ref{#1}}

\newcommand{\pinv}[1]{\ensuremath{#1^{\dagger}}} 	% Moore-Penrose pseudo-inverse

\DeclareMathOperator*{\rank}{rank}

\newcommand{\norm}[1]{\left\lVert#1\right\rVert}
\newcommand{\normsq}[1]{\left\lVert#1\right\rVert^2}
\newcommand{\frobenius}[1]{\left\lVert#1\right\rVert_\textrm{F}}
\newcommand{\frobeniusq}[1]{\left\lVert#1\right\rVert^2_\textrm{F}}
\newcommand{\abs}[1]{\left\lvert#1\right\rvert}
\newcommand{\absq}[1]{\left\lvert#1\right\rvert^2}
\newcommand{\Tr}[1]{\mathrm{Tr}\left(#1\right)}

\begin{document}

	\title{Cell-Free ISAC MIMO Systems: \\ Joint Sensing and Communication Beamforming}

	\author{Umut Demirhan and
		Ahmed Alkhateeb
		\thanks{Part of this work has been accepted in Asilomar Conference on on Signals, Systems, and Computers, 2023\cite{demirhan2023ISACmaxmin}. The authors are with the School of Electrical, Computer and Energy Engineering, Arizona State University, (Email: udemirhan, alkhateeb@asu.edu). This work is supported by the National Science Foundation under Grant No. 2229530. }}
	
	\maketitle
	
	\begin{abstract}		
		This paper considers a cell-free integrated sensing and communication (ISAC) MIMO system, where distributed MIMO access points (APs) jointly serve the communication users and sense the target. For this setup, we derive a sensing SNR for multi-static sensing where both joint communication and sensing signals transmitted by different APs are utilized. With this sensing objective, we develop two baseline approaches that separately design the sensing and communication beamforming vectors, namely communication-prioritized sensing beamforming and sensing-prioritized communication beamforming. Then, we consider the joint sensing and communication (JSC) beamforming design and derive the optimal structure of these beamforming vectors based on a max-min fairness formulation. In addition, considering any pre-determined JSC beam design, we devise a power allocation approach. The results show that the developed JSC beamforming is capable of achieving nearly the same communication signal-to-interference-plus-noise ratio (SINR) of the communication-prioritized sensing beamforming solution with almost the same sensing SNR of the sensing-prioritized communication beamforming approach. The proposed JSC beamforming optimization also provides a noticeable gain over the power allocation with regularized zero-forcing beamforming, yielding a promising strategy for cell-free ISAC MIMO systems. 
	\end{abstract}

	\section{Introduction}
	
	The integration of sensing functions into the communication systems is envisioned to be an integral part of the 6G and future communication systems \cite{liu2020joint6G, liu2022integrated, Demirhan_mgazine_radar, DeepSense}. If the hardware and wireless resources are efficiently shared, this can enable the communication infrastructure to have sensing capabilities at minimal cost and open the sensing frequency bands for wireless communication operation. The sensing capabilities may also be utilized to aid the communication system and improve its performance \cite{liu2020radar, demirhan2022beam, demirhan2022blockage, Digitaltwin}. They can also enable interesting applications in security, healthcare, and traffic management. Achieving efficient joint sensing and communication operation, however, requires the careful design of the various aspects of the integrated sensing and communication  (ISAC) system, including the transmission waveform, the post-processing of the received signals, and the MIMO beamforming. While these problems have recently attracted increasing research interest, the prior work has mainly focused on the single ISAC basestation case. In practice, however, multiple ISAC basestations will operate in the same geographical region, frequency band, and time, causing interference on each other for both the sensing and communication functions. This motivates the coordination between these distributed nodes to improve both communication and sensing performance. This ultimately leads to \textit{cell-free ISAC MIMO} systems, where distributed ISAC basestations jointly serve the same set of communication users and sense the same targets. With this motivation, this paper investigates the joint sensing and communication beamforming design of these cell-free ISAC MIMO systems.

	\subsection{Prior Work}
	Distributed antenna systems and interference management in the multi-cell MIMO networks have been extensively studied in the literature \cite{heath2011multiuser, schmidt2013comparison, zheng2015survey}. With the possibility of more extensive coordination among the basestations, coordinated multi-point transmissions \cite{nigam2014coordinated}, and, more recently, with the densification of the networks, cell-free massive MIMO \cite{ngo2017cell} have attracted significant interest. Cell-free massive MIMO is a concept where multiple access points (APs) jointly serve the user equipments (UEs) by transmitting messages to every user. Note that it is a distributed multi-user MIMO approach, and there are no limitations of cell boundaries. Due to its potential, various aspects of cell-free massive MIMO have been extensively investigated for further improvements \cite{nayebi2017precoding, ngo2017total, femenias2019cell, zhou2020max, bjornson2020scalable, demirhan2022enabling}. For example, precoding techniques for cell-free massive MIMO are studied in \cite{nayebi2017precoding}, energy minimization in \cite{ngo2017total}, fronthaul limitations in \cite{femenias2019cell}, scalability aspects in \cite{bjornson2020scalable}, and wireless fronthaul in \cite{demirhan2022enabling}. Most of these studies, however, did not include the unification of the sensing and communication functions in cell-free massive MIMO networks. 
	
	The literature for joint sensing and communication (JSC), also called dual-functional radar-communication (DFRC), has mainly focused on the single node (basestation) scenarios \cite{liu2018toward, fortunati2020massive, liu2020joint, cheng2021hybrid, bazzi2023outage}. For example, the design of the JSC waveform is studied in \cite{liu2018toward}. Specifically, the author investigated the JSC waveform design for correlated and uncorrelated waveforms and the trade-offs between communication and sensing. The authors in \cite{fortunati2020massive} proposed sensing post-processing for JSC systems. For beamforming, the work in \cite{liu2020joint} investigated the JSC beamforming design of a co-located MIMO system with monostatic radar that serves multiple users. The hybrid beamforming design for OFDM DFRC system is studied in \cite{cheng2021hybrid}. The optimal beamforming solution for JSC with and without sensing signal's successive interference cancellation is provided in \cite{hua2021optimal}. Along a similar direction, \cite{bazzi2023outage} formulated an outage-based beamforming problem and provided the optimal solution.     
	
	More relevantly, JSC with distributed nodes (basestations) has been investigated in a few papers \cite{huang2022coordinated, zhao2022joint, cheng2022coordinated, behdad2022power} for power allocation and beamforming. In most of them, however, each user is served by a single AP \cite{huang2022coordinated, zhao2022joint, cheng2022coordinated}; hence, these studies focused on the interference and not considered a fully cell-free MIMO setup. For example, \cite{zhao2022joint} proposes a transmit and receive beamforming optimization for JSC in a single basestation multi-user scenario, where the signals from a different cell are used to improve the sensing performance without communication interference. In \cite{huang2022coordinated}, a power allocation problem for JSC is formulated. The problem, however, assumes a single antenna system, and every UE is served by a single AP. In \cite{cheng2022coordinated}, the authors proposed a JSC beamforming optimization for maximizing the detection, however, this work only relied on each AP serving a single UE. As the most relevant work, the optimization of the JSC power allocation for cell-free massive MIMO has been investigated in \cite{behdad2022power}. The authors in this work adopted fixed beam designs, i.e., regularized zero beamforming for the communication with the sensing beamforming in the nullspace of the communication channels without further optimization, and focused on optimizing the power allocated to these beams. For the solution, they proposed a convex-concave procedure. Since these cell-free ISAC MIMO systems rely mainly on beamforming in their dual-function operation, it is very important to optimize the design of these JSC beams, which to the best of our knowledge, has not been previously investigated. With this motivation, we propose and compare various beamforming strategies for the cell-free ISAC MIMO systems.
	
	\subsection{Contributions}
	
	To investigate the JSC transmit beamforming in cell-free massive MIMO systems, in this paper, we consider a system model with many APs and UEs, where the APs jointly serve the UEs and sense the targets in the environment. With this model, we formulate beamforming and power allocation problems and develop various solutions. Our contributions in this paper can be summarized as follows:
	\begin{itemize}
		\item For beamforming, we first present two baseline strategies that we call \textit{communication-prioritized sensing} and \textit{sensing-prioritized communication beamforming}. In these strategies, either the sensing or the communication beamforming is given the priority to be designed first without accounting for the other function, and then the beamforming of the other function is designed in a way that does not affect the performance of the higher-priority function. For the communication-prioritized sensing, we develop a communication beamforming optimization approach for the given sensing beamforming vector. In this approach, we cast the problem as a convex second-order cone program and provide the optimal solution.
		\item We consider the case when the sensing and communication beamforming is jointly designed. For this, we formulate a JSC beamforming problem that aims to maximize the sensing SNR while satisfying the communication SINR constraints. We then re-formulate this problem as a non-convex semi-definite problem (SDP) and apply semi-definite relaxation (SDR) to find the optimal beamforming structure for a large set of classes.
		\item As an alternative to the joint beamforming design problem, we develop a power allocation problem for the given beamforming vectors. Specifically, with pre-determined beams, we reformulate our beamforming problem as a power allocation problem, which is also an SDP. For this, we develop an SDR-based solution.
	\end{itemize}
	%We have extensively evaluated the proposed approaches and showed that the JSC beamforming design provides near-optimal performance for both sensing and communication thanks to the co-design for the two functions. Our results highlight that the direct JSC beamforming optimization is a promising approach, enabling future cell-free communication and sensing systems.
	
	We have extensively evaluated the proposed approaches and showed that the JSC beamforming design provides better sensing performance than the sensing-prioritized solution while achieving the communication rates provided by the communication-prioritized solution. This is thanks to the co-design of the communication and sensing functions. Further, the JSC beamforming design overperforms the JSC power allocation for the regularized zero-forcing beams with a significant sensing SNR gain while providing the same communication rates. This shows the advantage of the joint beamforming optimization, making it desirable for future cell-free ISAC MIMO systems.
	
	\textbf{Organization}: In \sref{sec:systemmodel}, we present our system model with the communication and sensing objectives. In \sref{sec:comm-prioritized} and \ref{sec:sensing-prioritized}, we respectively present the communication- and sensing-prioritized beamforming approaches. Then, we develop the joint sensing and communication beamforming optimization in \sref{sec:beamforming}, and the power allocation formulation and solution in \sref{sec:power-allocation}. Finally, in \sref{sec:results}, we provide the numerical results evaluating the developed solutions and present our conclusions of the paper in \sref{sec:conclusion}.

	\textbf{Notation}: We use the following notation throughout this paper: $\bA$ is a matrix, $\ba$ is a vector, $a$ is a scalar, $\cA$ is a set. $\bA^T$, $\bA^H$, $\bA^*$, $\bA^{-1}$, $\pinv{\bA}$ are transpose, Hermitian (conjugate transpose), conjugate, inverse, and pseudo-inverse of $\bA$, respectively. $\norm{\ba}$ is the $l_2$-norm of $\ba$ and $\frobenius{\bA}$ is the Frobenius norms of $\bA$. $\bI$. $\mathcal{CN}(\bm{\mu},\bm{\Sigma})$ is a complex Gaussian random vector with mean $\bm{\mu}$ and covariance $\bm{\Sigma}$. $\bbE\left[\cdot\right]$ and $\otimes$ denote expectation and Kronecker product, respectively. $\mathbb{S}^+$ is the set of hermitian positive semidefinite matrices.
	
	\begin{figure}[!t]
		\centering
		\includegraphics[width=1\linewidth]{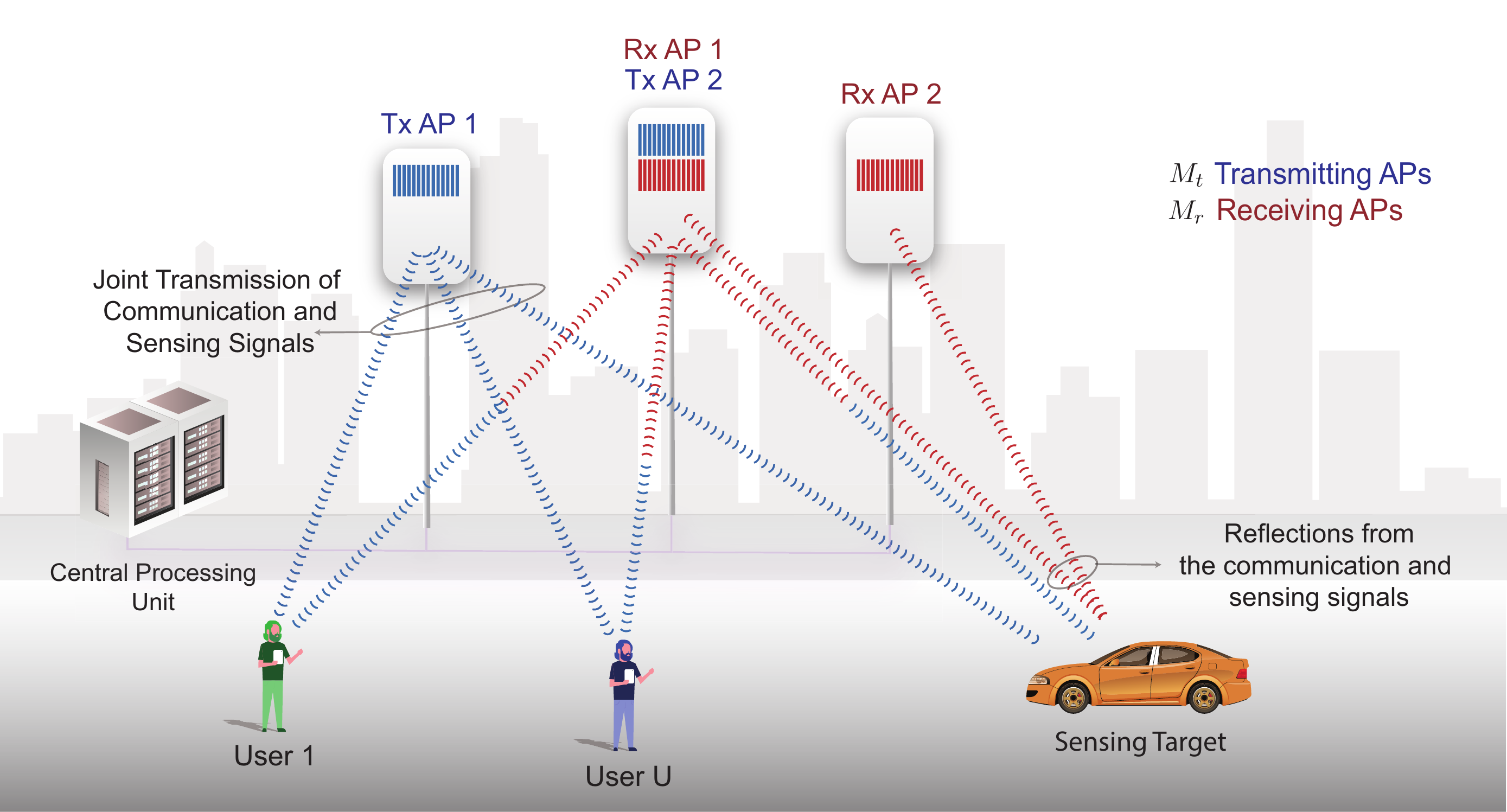}
		\caption{The system model with the joint sensing and communication transmissions is illustrated. The APs serve multiple users while aiming to sense the target.}
		\label{fig:systemmodel}
	\end{figure}
	
	\section{System Model} \label{sec:systemmodel}
	
	We consider a cell-free massive MIMO ISAC system with $M$ access points (APs)  and $U$ communication users, as illustrated in \figref{fig:systemmodel}. In the downlink, and without loss of generality, we assume that a subset $\mathcal{M}_t$ (out of the $M$ APs) are transmitting communication and sensing waveforms to jointly serve the $U$ users, where  $\left|\mathcal{M}_t\right|=M_t$. Simultaneously, a subset $\mathcal{M}_r$ (out of the $M$ APs) is receiving the possible reflections/scattering of the transmitted waveforms on the various targets/objects in the environment, with $\left|\mathcal{M}_r\right|=M_r$. It is important to note here that the subsets $\mathcal{M}_t$ and $\mathcal{M}_r$  may generally have no, partial, or full overlap, which means that none, some, or all the APs could be part of $\mathcal{M}_t$ and $\mathcal{M}_r$ and are simultaneously transmitting and receiving signals.  
	The transmitting and receiving APs are equipped with $N_t$ and $N_r$ antennas. Further, for simplicity, all the APs are assumed to have digital beamforming capabilities, i.e., each antenna element has a dedicated radio frequency (RF) chain. The UEs are equipped with single antennas. The APs are connected to a central unit that allows joint design and processing, and they are assumed to be fully synchronized for both sensing and communication purposes. 
	
	\subsection{Signal Model}
	In this subsection, we define the joint sensing and communication signal model for the downlink transmissions. The APs jointly transmit $U$  communication streams, $\{x_u[\ell]\}_{u\in\mathcal{U}}$, and $Q$ sensing streams, $\{x_q[\ell]\}_{q\in\mathcal{Q}}$, where $\mathcal{Q}=\{U+1, \ldots, U+Q\}$ and with $\ell$ denoting the $\ell$'s symbol in these communication/sensing streams. For ease of exposition, we also define the overall set of streams as $\mathcal{S}=\mathcal{U}\cup\mathcal{Q}=\{1, \ldots, S\}$ with $S=U+Q$. If $\bx_{m}[\ell] \in \mathbb{C}^{N_t \times 1}$ denotes the transmit signal from the transmitting AP $m$ due to the $\ell$-th symbol, we can then write
	\begin{equation} \label{eq:transmitsymbol}
		\bx_{m}[\ell] = \underbrace{\sum_{u \in \, \mathcal{U} }\bff_{m u} x_u[\ell]}_{\textrm{Communication}} + \underbrace{\sum_{q \in \mathcal{Q}} \bff_{m q} x_q[\ell]}_{\textrm{Sensing}} = \sum_{s \in \mathcal{S}} \bff_{m s} x_s[\ell],
	\end{equation}
	where $x_s[\ell] \in\mathbb{C}$ is the $\ell$-th symbol of the $s$-th stream, $\bff_{m s} \in \mathbb{C}^{N_t \times 1}$ is the beamforming vector for this stream applied by AP $m$. The symbols are assumed to be of unit average energy, $\bE[|x_s|^2]=1$. The beamforming vectors are subject to the total power constraint, $P_{m}$, given as
	\begin{equation}
		\bE[\normsq{\bx_{m}[\ell]}] = \sum_{s\in\mathcal{S}}\normsq{\bff_{m s}} \leq P_{m}.
	\end{equation}
	Further, by stacking the beamforming vectors of stream s of all the APs, we define the  beamforming vector $\bff_s $
	\begin{equation}\label{eq:beamvector}
		\bff_s = \begin{bmatrix} \bff^T_{1s} & \ldots & \bff_{M_t s}^T \end{bmatrix}^T \in \mathbb{C}^{M_t N_t}. 
	\end{equation}
	For each stream $s$, we denote the sequence of $L$ transmit symbols as $\bx_s = \begin{bmatrix} x_s[1], \ldots, x_s[L] \end{bmatrix}^T$. Given this notation, we make the following assumption, which is commonly adopted in the literature \cite{liu2020joint}: The messages of the radar and communication signals are statistically independent, i.e., $\bE[\bx_s \bx_s^H] = \bI$ and $\bE[\bx_s \bx_{s'}^H]=\bm{0}$ for $s, s' \in \mathcal{S}$ with $s \neq s'$. Note that the radar signal generation with these properties may be achieved through pseudo-random coding \cite{liu2020joint}.

	\subsection{Communication Model}
	We denote the communication channel between UE $u$ and AP $m$ as $\bh_{mu} \in \mathbb{C}^{N_t \times 1}$. Further, by stacking the channels between user $u$ and all the APs, we construct $\bh_{u} \in \mathbb{C}^{M_t N_t \times 1}$. Next, considering a block fading channel model, where the channel remains constant over the transmission of the $L$ symbols, we can write the received signal at UE $u$ as
	\begin{align}
		\begin{split}
			y^\textrm{(c)}_u[\ell] &= \sum_{m \in \mathcal{M}_t} \bh^H_{mu} \bx_m[\ell] + n_u \\
			&= \underbrace{\sum_{m \in \mathcal{M}_t} \bh^H_{mu} \bff_{mu} x_u[\ell]}_{\textrm{Desired Signal (DS)}} + \underbrace{\sum_{u' \in \mathcal{U} \backslash \{u\}} \sum_{m \in \mathcal{M}_t} \bh^H_{mu}  \bff_{mu'} x_{u'}[\ell]}_\textrm{Multi-user Interference (MUI)} 
			\\
			&\quad + \underbrace{\sum_{q\in\mathcal{Q}} \sum_{m \in \mathcal{M}_t} \bh^H_{mu} \bff_{mq} x_q[\ell]}_\textrm{Sensing Interference (SI)} + \underbrace{n_u[\ell]}_{\textrm{Noise}},
		\end{split}
		\label{eq:commsignal}
	\end{align}
	where $n_u[\ell] \sim \mathcal{CN}(0, \sigma_{u}^2)$ is the receiver noise of UE $u$. Then, the communication SINR of UE $u$ can be obtained as
	\begin{equation}
		\textrm{SINR}^\textrm{(c)}_u = \frac{\bbE[\absq{\textrm{DS}}]}{\bbE[\absq{\textrm{MUI}}] + \bbE[\absq{\textrm{SI}}] + \bbE[\absq{\textrm{Noise}}] }.
	\end{equation}
	In terms of the individual beamforming variables, $\{\bff_{ms}\}$, this SINR can be written as in \eqref{eq:SINR-cf}.
	\begin{figure*}[t]
		\begin{equation}  \label{eq:SINR-cf}
			\textrm{SINR}^\textrm{(c)}_u = \frac{\absq{\sum_{m \in \mathcal{M}_t} \bh^H_{mu} \bff_{mu}}}{\sum_{u' \in \mathcal{U} \backslash \{u\}} \absq{\sum_{m \in \mathcal{M}_t} \bh^H_{mu}  \bff_{mu'}} +  \sum_{q \in \mathcal{Q}} \absq{ \sum_{m \in \mathcal{M}_t} \bh^H_{mu} \bff_{mq}} + \sigma_{u}^2},
		\end{equation}
	\end{figure*}
	Further, we also write this expression in terms of the stacked vector variables of each UE $u$ as in \eqref{eq:beamvector}, as
	\begin{equation}  \label{eq:SINR}
		\textrm{SINR}^\textrm{(c)}_u = \frac{\absq{ \bh^H_{u} \bff_{u}}}{\sum_{u' \in \mathcal{U} \backslash \{u\}} \absq{\bh^H_{u}  \bff_{u'}} +  \sum_{q \in \mathcal{Q}} \absq{ \bh^H_{u} \bff_{q}} + \sigma_{u}^2}.
	\end{equation}
	
	%Note that \eqref{eq:SINR} simplifies the summation inside the absolute terms, which allows formulating the beamforming and power allocation optimization formulations in \sref{sec:beamforming} and \ref{sec:power-allocation}.

	\subsection{Sensing Model}
	For the sensing channel model, we consider a single-point reflector, as commonly adopted in the literature \cite{behdad2022power, Liu2022crlb}. Specifically, the transmit signal is scattered from the single-point reflector and received by the receiving APs in $\mathcal{M}_r$. With a single path model, the channel between the transmitting AP $m_t$ and the receiving AP $m_r$ through the reflector is defined as
	\begin{align}
		\begin{split}
			\bG_{m_t m_r} &= {\alpha}_{m_t m_r} \ba(\theta_{m_r}) \ba^H(\theta_{m_t}),
		\end{split}
	\end{align}
	where $\alpha_{m_t m_r} \sim \mathcal{CN}(0, \zeta_{m_t m_r}^2)$ is the combined sensing channel gain, which includes the effects due to the path-loss and radar cross section (RCS) of the target. $\ba(\theta)$ is the array response vector. The angles of departure/arrival of the transmitting AP $m_t$ and receiving AP $m_r$ from the point reflector are respectively denoted by $\theta_{m_t}$ and  $\theta_{m_r}$. We consider the Swerling-I model for the sensing channel \cite{richards2010principles}, which assumes that the fluctuations of RCS are slow and the sensing channel does not change within the transmission of the $L$ sensing and communication symbols in $\bx_s$. With this model, the signal received at AP $m_r$ at instant $\ell$ can be written as
	\begin{align} \label{eq:radar_rx}
		\begin{split}
			\by^\textrm{(s)}_{m_r}[\ell] &= \sum_{m_t\in \mathcal{M}_t} \bG_{m_t m_r} \, \bx_{m_t}[\ell] + \bn_{m_r}[\ell] \\
			&= \sum_{m_t\in \mathcal{M}_t} {\alpha}_{m_t m_r} \ba(\theta_{m_r}) \ba^H(\theta_{m_t}) \, \bx_{m_t}[\ell] + \bn_{m_r}[\ell],
		\end{split}
	\end{align}
	where $\bn_{m_r}[\ell]\in\mathbb{C}^{N_r}$ is the receiver noise at AP $m_r$ and has the distribution $\mathcal{CN}(0, \varsigma_{m_r}^2 \bI)$. To write the received radar signal due to the $L$ symbols in a compact form, we introduce
	\begin{align}
		\overline{\bF}_m &= [\bff_{m1}, \ldots, \bff_{mS}] \in \mathbb{C}^{N_t \times S}, \\
		\overline{\bX} &= [\bx_{1}, \ldots, \bx_{S}]^T \in \mathbb{C}^{S \times L}.
	\end{align}
	Then, we can write the transmit signal from each AP $m_t$, in \eqref{eq:transmitsymbol}, due to the $L$ symbols as
	\begin{equation}
		\bX_{m_t} = \overline{\bF}_{m_t} \overline{\bX} \in \mathbb{C}^{N_t \times L}.
	\end{equation}
	With that, we can re-write the sensing signal in \eqref{eq:radar_rx} at each receiving AP $m_r$, due to the $L$ symbols, in a compact form as
	\begin{align} \label{eq:simplified_form}
		\begin{split}
			\bY^\textrm{(s)}_{m_r} &= \underbrace{\sum_{m_t\in \mathcal{M}} \underbrace{ {\alpha}_{m_t m_r} \ba(\theta_{m_r}) \ba^H(\theta_{m_t}) \overline{\bF}_{m_t}}_{\triangleq \overline\bG_{m_t m_r}}}_{\triangleq \overline\bG_{m_r}} \overline{\bX} + \bN_{m_r},
		\end{split}
	\end{align}
	with $\overline{\bG}_{m_r}$ denoting the beam-space sensing channel of the receiving AP $m_r$ and the receive noise matrix $\bN_{m_r} = [\bn_{m_r}[1], \ldots, \bn_{m_r}[L]]$. 
	
	To define a general sensing objective that is correlated with the performance of various sensing tasks (e.g., detection, range/Doppler/angle estimation and tracking), we adopt the joint SNR of the received signals as the sensing objective. Note that the use of the joint SNR requires joint processing of the radar signal at the $\mathcal{M}_r$ sensing receivers. The sensing SNR can be written as
	\begin{align} \label{eq:SNRsensing}
		\begin{split}
			\textrm{SNR}^\textrm{(s)} &= \frac{ \bbE\left[\sum_{m_r \in \mathcal{M}_r}  \frobeniusq{ \overline\bG_{m_r} \overline\bX}\right] }{\bbE\left[ \sum_{m_r \in \mathcal{M}_r} \frobeniusq{\bN_{m_r}}\right]} \\
			&= \frac{ \sum_{m_r \in \mathcal{M}_r}  \sum_{m_t\in \mathcal{M}_t} \zeta_{m_t m_r} ^2  \normsq{\ba^H(\theta_{m_t}) \overline{\bF}_{m_t}} }{ \sum_{m_r \in \mathcal{M}_r} \varsigma_{m_r}^2},
		\end{split}
	\end{align}
	where the derivation is provided in Appendix \ref{appendix-sensingSNR}. Recall that $\zeta_{m_t m_r}^2$ denotes the variance of the combined sensing channel gain and $\varsigma_{m_r}^2$ is the variance of the radar receiver noise. The sensing SNR is scaled with the contribution of all the \textit{communication and sensing} streams.

	\textbf{Our objective} is then to design the cell-free communication beamforming $\left\{\bff_u\right\}_{u \in \mathcal{U}}$ and the sensing beamforming $\left\{\bff_q\right\}_{q \in \mathcal{Q}}$ to optimize the communication SINR and the sensing SNR defined in \eqref{eq:SINR} and \eqref{eq:SNRsensing}. It is important to note here that, in this paper, we focus on the beamforming design assuming that the communication channel and the sensing target angles are known to the transmit APs. Extending this work to include imperfect channels/angle knowledge is an interesting direction. In the next three sections, we present the proposed beamforming strategies for communication-prioritized sensing, sensing-prioritized communication, and joint sensing and communication.
	
	%In the following, we assume that the transmitting APs have the knowledge of the target direction $\theta_m$.	
	
	%%%%%%%%%%%%%%%%%%%%%%%%%%%%%%%%%%%%%%%%%%%%%%%%
	\section{Communication-Prioritized Sensing Beamforming Design} \label{sec:comm-prioritized}
	In this section, we investigate the scenario where the communication has a higher priority, and where the communication beams are already designed a priori. In this case, the objective is to design the sensing beams to optimize the sensing performance while not affecting the communication performance (i.e., not causing any interference to the $U$ communication users).  Note that in this section and the next section, \sref{sec:sensing-prioritized}, we assume that $Q=1$ since we have one sensing target and that the total power is divided with a fixed ratio $\rho$, leading to $P^\textrm{c}_m = \rho P_m $ for the communication power and  $P^\textrm{s}_m = P_m - P^\textrm{c}_m$ for the sensing power. This makes it interesting to explore the joint optimization of the beamforming and power allocation in cell-free ISAC MIMO systems, which is presented in \sref{sec:power-allocation}. Next, we present two sensing beamforming design solutions for the cases (i) when the communication users are not present and when (ii) they are present.

	\textbf{Conjugate Sensing Beamforming:} When the communication users are not present (i.e., $U=0$), for example, during downtimes, the system can completely focus on the sensing function. In this case, and given the single target sensing model, the conjugate sensing beamforming solution becomes optimal, as it directly maximizes the sensing SNR.  With this solution, the sensing beamforming vectors can be written as 
	\begin{equation} \label{eq:sensing-CB}
		\bff_{m q}^\mathrm{CB} = \sqrt{\frac{p_{mq}}{N_t}} \, \ba(\theta_{m}),
	\end{equation}
	where $p_{mq} = P^s_m$ is the power allocated for the sensing beam.  
	
	\textbf{Communication-Prioritized Optimal Sensing Solution:} When the communication users exist (i.e., $U \geq 1$), and since the communication has a higher priority, a straight-forward optimal sensing beamforming approach is to project the optimal sensing beams (constructed through conjugate beamforming) to the null-space of the communication channels. This way, the interference contribution of the sensing beam to the communication channels is eliminated while the sensing SNR is maximized within the communication null space. Let $\bH_{m}=[\bh_{m1}, \ldots, \bh_{mU}] \in \mathbb{C}^{N_t \times U}$ denote the full channel matrix from the transmit AP $m$ to all the UEs, then the NS sensing beamforming can be constructed as
	\begin{equation} \label{eq:sensing-NS}
		\bff_{m q}^\mathrm{NS} = \sqrt{p_{mq}} \frac{\left(\bI - \bH_m \left( \bH_{m}^H \bH_m\right)^\dagger \bH_m^H \right) \ba(\theta_{m})}{\norm{\left(\bI - \bH_m \left( \bH_{m}^H \bH_m\right)^\dagger \bH_m^H \right) \ba(\theta_{m})}},
	\end{equation}
	where we again set the allocated power $p_{mq}=P^s_m$ as we consider a single sensing beam.
	
	\section{Sensing-Prioritized Communication Beamforming Design} \label{sec:sensing-prioritized}
	In this section, we consider the scenario where the sensing has a higher priority, and where the sensing beams are already designed a priori. In this case, the objective is to design the communication beams to optimize the communication performance while \textit{minimizing} the impact of the sensing interference. It is important to note here that an interesting difference between the communication and sensing optimization problems is that while the sensing signals could cause interference that degrades the communication performance, the communication signals could generally be leveraged to further enhance the sensing performance. Next, we present two communication beamforming design solutions for the cases when (i) the sensing target is not present and when (ii) it is present.

	\textbf{Regularized Zero-forcing Beamforming:} When the sensing target is not present, i.e., $Q=0$, a near-optimal communication beamforming design is the regularized zero-forcing (RZF) \cite{bjornson2014optimal}. This solution allows a trade-off between the multi-user interference and noise terms of the SINR through a regularization parameter $\lambda$, that is added to the ZF beamforming:
	\begin{equation}\label{eq:comm-RZF}
		\tilde{\bff}_u^\mathrm{RZF} =  \left(\lambda \bI + \sum_{u'\in\mathcal{U}} \bh_{u'} \bh_{u'}^H \right)^{-1} \bh_u,
	\end{equation}
	which then can be normalized to satisfy the power constraints, i.e., $\bff_{mu}^\mathrm{RZF} = \sqrt{p_{mu}} ({\tilde{\bff}^\mathrm{RZF}_{mu}}/{|\tilde{\bff}^\mathrm{RZF}_{mu}|})$. We here again adopt $p_{mu}=P_m/U$  with an equal power between the beams. For the RZF, it is preferable to have a higher regularization parameter in the scenarios with higher noise, and smaller in scenarios with more interference.  For further details, we refer to \cite{bjornson2014optimal}.

	\textbf{Sensing-Prioritized Optimal Communication Solution:}  For the case when the sensing beam is designed a priori, we derive a max-min fair rate optimal communication beamforming solution. First, this max-min problem can be written as 
	\begin{subequations} \label{eq:maxminSINR}
		\begin{align}
			\textrm{(P1.1):}\quad\quad \max_{\{\bff_{mu}\}} \min_u\quad & \textrm{SINR}_u^\mathrm{(c)}
			\\
			\mathrm{s.t.}\quad & \sum_{u\in\mathcal{U}}\normsq{\bff_{mu}} \leq P^c_m, \quad \forall m \in \mathcal{M}_t,
		\end{align}
	\end{subequations}
	where the objective is quasiconvex \cite{boyd2004convex} and shows a similar structure to the optimal beamforming formulation for the cell-free massive MIMO networks with only the communication objective \cite{zhou2020max}. For a given minimum SINR constraint $\gamma$, (P1.1) can be written as the feasibility problem
	\begin{subequations} \label{eq:SOCP}
		\begin{align}  \label{eq:SOCPoriginal}
			\textrm{(P1.2):}\quad\quad \mathrm{find} \quad &{\{\bff_{mu}\}}
			\\
			\mathrm{s.t.}\quad & \textrm{SINR}_u^\mathrm{(c)} \geq \gamma, \quad \forall u \in \mathcal{U}, \label{eq:SINRconstraint}
			\\
			&\sum_{u\in\mathcal{U}}\normsq{\bff_{mu}} \leq P^c_m, \quad \forall m \in \mathcal{M}_t.
		\end{align}
	\end{subequations}
	
	Here, we note that the SINR constraint \eqref{eq:SINRconstraint} is in a fractional form. This, however, can be converted to a second-order cone constraint. For this purpose, we can re-write the constraint as $\left(1+\frac{1}{\gamma}\right)\absq{\sum_{m \in \mathcal{M}_t} \bh^H_{mu} \bff_{mu}} \geq \sum_{u' \in \mathcal{U}} \absq{\sum_{m \in \mathcal{M}_t} \bh^H_{mu}  \bff_{mu'}} + \sum_{q \in \mathcal{Q}} \absq{ \sum_{m \in \mathcal{M}_t} \bh^H_{mu} \bff_{mq}} + \sigma_{u}^2$.
	Now, taking the square root of both sides, we can convert the given form to a second-order cone constraint. The square root, however, leaves an absolute on the left-hand side, which is a non-linear function. This can be simplified as the real part of the variable \cite{bjornson2014optimal}, since any angular rotation ($e^{-j \psi}$) to the expression inside the absolute does not change the value, i.e.,
	$\abs{\sum_{m \in \mathcal{M}_t} \bh^H_{mu} \bff_{mu}} = \abs{\sum_{m \in \mathcal{M}_t} \bh^H_{mu} \bff_{mu} e^{-j\psi}} = \textrm{Re}\left\{\sum_{m \in \mathcal{M}_t} \bh^H_{mu} \bff_{mu}\right\}$. This approach can be seen as selecting the optimal solution with a specific angular rotation from the set of infinite rotations $\psi\in[0, 2\pi)$. Finally, we can write the constraint \eqref{eq:SINRconstraint} as a second-order cone as follows
	\begin{align} \label{eq:SOCPfinal}
		\left(1+\frac{1}{\gamma}\right)^{\frac{1}{2}} \, \textrm{Re}\left\{\sum_{m \in \mathcal{M}_t} \bh^H_{mu} \bff_{mu}\right\} \geq
		\begin{Vmatrix}
			\sum_{m \in \mathcal{M}_t} \bh^H_{mu}  \bff_{m1}\\
			\vdots\\
			\sum_{m \in \mathcal{M}_t} \bh^H_{mu} \bff_{mS} \\
			\sigma_{u}
		\end{Vmatrix}.
	\end{align}
	When \eqref{eq:SINRconstraint} is replaced with  \eqref{eq:SOCPfinal}, it results in a second-order cone problem and can be solved by the convex solvers \cite{CVX}. Using the bisection algorithm, the maximum SINR value, $\gamma^\star$, can be obtained by solving the convex feasibility problem \eqref{eq:SOCP} for different values of $\gamma$ within a predetermined range $[\gamma_\textrm{min}, \gamma_\textrm{max}]$. This computes the optimal solution to  \eqref{eq:maxminSINR}.
	
	%%%%%%%%%%%%%%%%%%%%%%%%%%%%%%%%%%%%%%%%%%%%%%%%
	\section{Joint Sensing and Communication: Beamforming Optimization} \label{sec:beamforming}
	A more desirable approach for cell-free joint sensing and communication MIMO systems  is to jointly optimize the beamforming vectors for the sensing and communication functions. Specifically, our objective is to maximize the sensing SNR together with the communication SINR of the UEs. Towards this objective, we reformulate \eqref{eq:SOCP} as a sensing SNR maximization problem by (i) adding the sensing SNR maximization as an objective to the feasibility problem, and (ii) generalizing the minimum communication SINR limit, $\gamma$, individually for each UE with $\gamma_u$. Then, the JSC beamforming optimization problem can be written as
	\begin{subequations} \label{eq:JSC}
		\begin{align}
			\textrm{(P2.1):}\quad\quad \max_{\{\bff_{ms}\}} \quad &\textrm{SNR}^\textrm{(s)}
			\\
			\mathrm{s.t.}\quad & \textrm{SINR}^\mathrm{(c)}_u \geq \gamma_u, \quad \forall u \in \mathcal{U}, \label{eq:SINRconstraintJSC}
			\\
			&\sum_{s\in\mathcal{S}}\normsq{\bff_{ms}} \leq P_m, \quad \forall m \in \mathcal{M}_t, \label{eq:powerconstraintJSC}
		\end{align}
	\end{subequations}
	where the objective, i.e., the maximization of the convex SNR expression, $\textrm{SNR}^\textrm{(s)}$, is non-convex and the problem is a non-convex quadratically constrained quadratic program (QCQP). Hence, a similar approach to the beamforming optimization in the previous section can not be adopted. The problem in \eqref{eq:JSC}, however, can be cast as a semidefinite program, which allows applying a semidefinite relaxation for the non-convex objective \cite{luo2010semidefinite}. With the relaxation, the problem becomes convex, and the optimal solution can be obtained with the convex solvers. After that, the solution from the relaxed problem can be cast to the original problem's space with a method designed specifically for the problem. In the following, we present the details of our approach.

	To reformulate \eqref{eq:JSC} as an SDP, we first re-define the beamforming optimization variables as matrices: $\bF_s = \bff_s \bff_s^H$, $\forall s \in \mathcal{S}$. Writing \eqref{eq:JSC} in terms of $\bF_s$ instead of $\bff_s$ eliminates the quadratic terms in the sensing SNR and communication SINR expressions. 
	This SDP formulation, however, by construction introduces two new constraints: (i) The convex hermitian positive semi-definiteness constraint $\bF_s \in \mathbb{S}^+$, where $ \mathbb{S}^+$ is the set of hermitian positive semidefinite matrices, and (ii) the non-convex rank-1 constraint $\mathrm{rank}(\bF_s)=1$. 
	Further, we need to write the problem (P2.1) in terms of these newly introduced variables, $\{\bF_s\}$. For this purpose, we define the AP selection matrix, $\bD_{m} \in \mathbb{R}^{MN_t \times M N_t}$, where each element of this matrix is given by
	\begin{equation}
		[\bD_{m}]_{ij} = 
		\begin{cases}
			1& \textrm{if $(m-1) N_t +1 \leq i\leq m N_t$ with $i=j$}, \\
			0& \textrm{otherwise}.
		\end{cases}
	\end{equation}
	where the only non-zero elements of the $\bD_m$ is the identity matrix placed at the $m$-th cross diagonal $N_t \times N_t$ block matrix.
	To write the sensing SNR in a compact form, we define  $\bA=\sum_{m_t\in \mathcal{M}_t} \bar{\zeta}_{m_t} \bD_{m_t},  \overline\bA \bD_{m_t}$, where $\overline\bA = \overline{\ba} \, \overline{\ba}^H$ with $\overline{\ba} = [\ba(\theta_1)^T, \ldots, \ba(\theta_{M_t})^T]^T$, and $\bar{\zeta}_{m_t}=\sum_{m_r \in \mathcal{M}_r} \zeta_{m_t, m_r}^2$. Now, we can write the objective of \eqref{eq:JSC} (sensing SNR) in terms of $\bA$ as 
	\begin{align}\label{eq:SNRsensingSDP}
		\textrm{SNR}^\textrm{(s)} &=
		\frac{ \Tr{ \bA \sum_{s  \in \mathcal{S}} \bF_{s} } }{\sum\limits_{m_r \in \mathcal{M}_r} \varsigma_{m_r}^2}.
	\end{align}
	where the derivation is provided in Appendix \ref{appendix-sensingSNR-SDP}.
	
	For the constraints of the problem in \eqref{eq:JSC}, we define $\bQ_u = \bh_u \bh_u^H$ and re-write the SINR in \eqref{eq:SINR} in terms of the new variables as
	\begin{align} \label{eq:SINRcommSDP}
		\textrm{SINR}_u^\mathrm{(c)} &= \frac{\Tr{\bQ_{u} \bF_u }}{\sum\limits_{u' \in \mathcal{U} \backslash \{u\}} \Tr{ \bQ_{u}  \bF_{u'}} +\sum\limits_{q\in \mathcal{Q}} \Tr{\bQ_{u} \bF_q} + \sigma_{u}^2}.
	\end{align}
	
	With this, we can write the constraint in \eqref{eq:SINRconstraintJSC} {and} the power constraint {in} \eqref{eq:powerconstraintJSC}  as
	\begin{align} \label{eq:SINRcommconstraintSDP}
		\begin{split}
			\left(1+\gamma_u^{-1}\right) \Tr{\bQ_{u} \bF_u } - \Tr{\bQ_{u} \sum_{s\in \mathcal{S}} \bF_s} \geq \sigma_{u}^2,
		\end{split}
	\end{align}
	\begin{equation} \label{eq:powerconstraintJSCSDP}
		\sum_{s\in\mathcal{S}}\Tr{\bD_m \bF_{s}} \leq P_m, \quad \forall m \in \mathcal{M}_t.
	\end{equation}
	
	Here, we notice that for the objective and constraints, we can simplify the sensing variables by defining $\bF_\mathcal{Q}=\sum_{q\in\mathcal{Q}} \bF_q$, since the sensing variables $\bF_q$ only appear in the defined summation form. For the optimality of the problem, however, this variable needs to have at most rank $Q$, so that we can construct $Q$ beamforming vectors\footnote{If the rank of this variable is less than $Q$, some of the beamforming vectors are not needed, and can be selected as zero.}.  Then, by collecting the expressions together, we can write the SDP form of our problem (P2-QCQP) as
	\begin{subequations} \label{eq:JSCSDP}
		\begin{align} 
			\textrm{(P2.1-SDP):}\quad \max_{\{\bF_{u}\}, \bF_\mathcal{Q}} \quad &  { \Tr{ \bA \sum_{s  \in \mathcal{S}} \bF_{s} } } \label{eq:JSCSDP-1}
			\\
			\mathrm{s.t.}\quad & \eqref{eq:SINRcommconstraintSDP} \textrm{ and } \eqref{eq:powerconstraintJSCSDP},
			\\
			& \bF_u \in \mathbb{S}^+ \ \ \forall u \in \mathcal{U}, \quad \bF_\mathcal{Q} \in \mathbb{S}^+ \label{eq:JSCSDP-3}
			\\
			& \mathrm{rank}(\bF_u)=1 \ \ \forall u \in \mathcal{U}, \label{eq:JSCSDP-4}\\
			& \mathrm{rank}(\bF_\mathcal{Q})\leq Q, \label{eq:JSCSDP-5}
		\end{align}
	\end{subequations}
	which can be relaxed by removing the rank constraints, i.e., \eqref{eq:JSCSDP-4}-\eqref{eq:JSCSDP-5}. This relaxed problem \textbf{(P2.1-SDR)}, defined as \eqref{eq:JSCSDP-1}-\eqref{eq:JSCSDP-3}, can be solved via CVX and convex SDP solvers \cite{CVX,SDPT3}. Then, if the matrices obtained by this solution, denoted by $\{\bF'_u\}$, are rank-1, and $\bF'_\mathcal{Q}$ is at most rank-$Q$, then they are optimal for \eqref{eq:JSCSDP}. The optimal user beamforming vectors, $\bff_u$, in this case, can be obtained as the eigenvector of $\bF'_u$. Similarly, $\{\bff_q\}$ can be constructed as the $Q$ eigenvectors of $\bF'_\mathcal{Q}$. For the case the user matrices are not rank-1, we make the following proposition.\\

	\noindent \textbf{Proposition 1.} \textit{There exists a solution to the problem \eqref{eq:JSCSDP}, denoted by $\{\bF''_u\}$ and $\bF''_\mathcal{Q}$, that satisfies $\mathrm{rank}(\bF''_u)=1$, $\forall u \in \mathcal{U}$ and 
		\begin{equation} \label{eq:proposition-sensingbeams}
			\bF''_\mathcal{Q} = \bF'_\mathcal{Q} + \sum_{u\in\mathcal{U}} \bF'_u - \sum_{u\in\mathcal{U}} \bF''_u.
		\end{equation} 
		where $\bF'_\mathcal{Q}$ and $\{\bF'_u\}$ are the solutions of the SDP problem in (P2.1-SDR). 
		The communication beamforming vectors of this solution can be given as
		\begin{equation} \label{eq:proposition}
			\bff''_u=(\bh_u^H \bF'_u \bh_u)^{-\frac{1}{2}} \bF'_u \bh_u.\\
		\end{equation}
		Further, if $\mathrm{rank}(\bF''_\mathcal{Q}) \leq Q$, the optimal sensing beamforming vectors of this solution can be constructed by
		\begin{equation} \label{eq:sensingapprox}
			\bff''_q = \sqrt{\lambda_{q-U}} \, {\bu}_{q-U},
		\end{equation}
		with $\lambda_i$ and $\bu_i$ being the $i$-th largest eigenvalue of $\bF''_\mathcal{Q}$ and the corresponding eigenvector.
	}
	
	The proof extends the solution in \cite[Theorem~1]{liu2020joint}, which we provide in Appendix \ref{sec:appendixb}. For $\mathrm{rank}(\bF''_\mathcal{Q}) \leq Q$, the solution obtained is from Proposition 1 optimal. In the case $\mathrm{rank}(\bF''_\mathcal{Q}) > Q$, however, \eqref{eq:sensingapprox} will not lead to the optimal solution. We will examine the performance of this approximation in \figref{fig:jscnumue}. Next, we investigate the value of $Q$ required to satisfy the optimality.
	
	\subsection{How Many Sensing Streams Do We Need?}
	In the formulations of (P2.1-SDP) and (P2.1-SDR) in \sref{sec:beamforming}, the number of sensing streams is kept generic with the variable $Q$. In reality, however, it would be preferable to have as few as possible sensing streams. To that end, it is interesting to investigate how many sensing beams are needed to achieve optimal sensing performance. For this objective, we can further investigate (P2.1-SDR) to find the constraints on the optimal sensing solution. Specifically, we attempt to solve the problem (P2.1-SDR). Since this problem is convex, it satisfies the strong duality \cite{boyd2004convex, xu2014multiuser}. Then, we can derive the dual problem as
	\begin{align} \label{eq:JSCSDR-dual}
		\begin{split}
			\textrm{(D2.1-SDR): } \min_{\{\lambda_u\}, \{\nu_m\}} \quad &  \sum_m \nu_m P_m - \sum_u \lambda_u \sigma_{u}^2
			\\
			\mathrm{s.t.} \quad & \bB_u \preceq 0 \ \ \forall u \in \mathcal{U}, \quad \bB_\mathcal{Q} \preceq 0
		\end{split}
	\end{align}
	where $\{\lambda_u\geq0\}$, $\{\nu_m\geq0\}$ are the Lagrangian coefficients corresponding to the SINR and power constraints, respectively, and
	\begin{align}
		\bB_u &= \bA + \lambda_u \gamma_u^{-1} \bQ_u - \sum_{u'\in \mathcal{U}\backslash \{u\}}\lambda_{u'} \bQ_{u'} - \sum_m \nu_m \bD_m,\label{eq:appendix-bu} \\
		\bB_\mathcal{Q} &= \bA - \sum_{u'\in \mathcal{U}}\lambda_{u'} \bQ_{u'} - \sum_m \nu_m \bD_m. 
	\end{align}
	The derivation of the dual function is provided in Appendix \ref{sec:appendixc}. Further, we make the following remark on the definition of the new variables in the dual problem.
	
	\noindent\textbf{Remark 1.}
	From the definition of the new variables, $\bB_u$ and $\bB_\mathcal{Q}$, we also have the relation
	\begin{equation} \label{eq:b-c-relation}
		\bB_u =  \bB_\mathcal{Q} + \lambda_u (1+\gamma^{-1}_u) \bQ_u.
	\end{equation}
	Let us assume that there exists a feasible set of primal-dual optimal variables, i.e., $\{\bF_u^\star\}$, $\bF_\mathcal{Q}^\star$, $\{\lambda_u^\star\}$, $\{\eta_m^\star\}$, and the corresponding variables $\bB^\star_\mathcal{Q}$ and $\{\bB^\star_u\}$. With the Karush-Kuhn-Tucker (KKT) conditions \cite{boyd2004convex}, we obtain the complementary slackness for the semidefinite constraints
	\begin{equation} \label{eq:F_q_ns}
		\begin{split}
			\bB_u^\star \bF_u^\star = \bm{0}  \quad \textrm{and} \quad \bB_\mathcal{Q}^\star \bF_\mathcal{Q}^\star = \bm{0} 
		\end{split}
	\end{equation}
	which shows that $\bF_\mathcal{Q}^\star$ is in the nullspace of $\bB_\mathcal{Q}^\star$. We can further refine this condition with the following proposition.
	
	\noindent \textbf{Proposition 2.} \textit{The sensing beamforming matrix is in the nullspace of $\bA-\sum\nu^\star_m \bD_m$. In addition, it is in the nullspace of $\bQ_u$ for any user with $\lambda^\star_u>0$}.
	
	\noindent\textbf{Proof:}
	We have
	\begin{equation}
		\begin{split}
			\lambda^\star_u (1+\gamma^{-1}_u) \Tr{\bQ_u \bar{\bF}^\star_\mathcal{Q}}  &= \Tr{(\bB^\star_u-\bB^\star_\mathcal{Q})\bar{\bF}^\star_\mathcal{Q}}  
			\\
			&= \Tr{\bB^\star_u \bar{\bF}^\star_\mathcal{Q}} 
			\\
			&\leq \max_{\bar{\bF}_\mathcal{Q} \succeq 0} \Tr{\bB^\star_u \bar{\bF}_\mathcal{Q}} = 0,
		\end{split}
	\end{equation}
	where the first equality is due to definition (Remark 1), the second equality by the complementary slackness condition given in \eqref{eq:F_q_ns}, and the latter inequality due to the multiplication of the negative and positive semidefinite matrices.
	
	This proposition shows that a sensing matrix will be in the nullspace of stacked UE channels, $\bh_u$, for every UE $u$ that satisfies the SINR constraint at the equality. On the other hand, when the equality of the SINR constraint is not satisfied, we have $\lambda_u^\star=0$ due to the complementary slackness condition of the SINR constraint. This leads to $\bB_u^\star=\bB_\mathcal{Q}^\star$ as shown in Remark 1. Combined with the complementary slackness conditions in \eqref{eq:F_q_ns}, it results in $\bF^\star_u$ and $\bF^\star_\mathcal{Q}$ being in the same space defined as the nullspace of the channels of the UEs with $\lambda_u^\star>0$ and the sensing direction via $\bA-\sum \nu^\star_m \bD_m$. There is, however, no enforcement towards the direction of the user itself because $\lambda^\star_u(1+\gamma_u^{-1})\bQ_u = 0$. Then, this case is likely to appear only if there is sufficient SINR with the transmission towards the sensing direction from all the APs (e.g., the UE and target are at the same location). Hence, it is trivial and not of significant interest. With this observation, we focus on the case with $\lambda^\star_u>0$ for every UE. 
	
	In Proposition 2, we also have the nullspace of $\bA-\sum \nu^\star_m \bD_m$ to define the sensing matrix. We note that $\sum \nu^\star_m \bD_m$ is a diagonal matrix with the diagonal of each block having the same value $\nu^\star_m$.
	
	\noindent\textbf{Remark 2.}
	The sensing SNR matrix, $\bA$, is a block diagonal matrix of rank-1 blocks, i.e., 
	\begin{equation}
		\bA = \mathrm{diag}(\bar{\zeta}_{1}\ba(\theta_1)\ba^H(\theta_1), \ldots, \bar{\zeta}_{M_t}\ba(\theta_{M_t})\ba^H(\theta_{M_t})).
	\end{equation}
	This fact can be seen by the definition of $\bA=\sum_{m_t\in \mathcal{M}_t} \bar{\zeta}_{m_t} \bD_{m_t} \overline\bA \bD_{m_t}$, where the multiplication of a matrix from both sides with the selection matrix, $\bD_{m_t}$, results in the block diagonal of the selected entries. Further, with a slight abuse of notation, each block of the diagonal $\bA_{m_t} \triangleq \bar{\zeta}_{m_t}\ba(\theta_{m_t})\ba^H(\theta_{m_t})=\bD_{m_t}\bA\bD_{m_t}$ is a weighted outer product of an array response vector. Therefore, we have $\rank(\bA_{m_t})=1, \forall m_t \in \mathcal{M}_t$ and $\rank(\bA)=M_t$.
	
	With Remark 2, we can further refine our space as $\bA-\sum \nu^\star_m \bD_m = \sum(\bA_m-\nu^\star_m \bI_m)$, where each component, $\bA_m-\nu^\star_m \bI_m$, is a diagonal block. For $\nu^\star_m>0$, each of these components can have a single nullspace vector $\bff_{mq}=\ba(\theta_m)$ if $\nu^\star_m=\bar{\zeta}_m$. Note that this only constructs a part of the sensing matrix, not the full domain $\bff_q$. If $\nu^\star_m\neq\bar{\zeta}_m$, the component is full rank, and the nullspace is empty. In other words, the existence of sensing beams is independently determined at each AP based on $\nu^\star_m$ and only available if $\nu^\star_m=\bar{\zeta}_m$. For any other $\nu^\star_m>0$, there is no sensing stream. Then, the sensing beams can be written in the form of $\bff_q=[\eta_{1q}\ba(\theta_{1}), \ldots, \eta_{Mq}\ba(\theta_{M})]$ for some $\eta_{mq} \in \mathbb{C}$, and this vector is in the nullspace of $\bh_u$ for every UE with $\lambda^\star_u>0$. To that end, further investigating the performance of the nullspace sensing beam with the suboptimal beamforming solutions is interesting. As an alternative to beamforming optimization, in the next section, we develop a power allocation approach for the pre-determined beamforming vectors. Before moving on, we further conclude the limitations on the sensing streams.
	
	\noindent \textbf{Proposition 3.} \textit{For $\nu_m^\star>0$ $\forall m \in \mathcal{M}_t$, the maximum number of sensing streams is limited by}
	\begin{equation}
		\rank(\bF^\star_\mathcal{Q})\leq M_t.
	\end{equation}
	The proof follows Remark 2, where the minimum rank of $\bA-\sum_m\nu ^\star_m\bD_m=(M_t-1)N_t$, and its nullspace can have at most $M_t$ dimensions. Further, we can refine the limit in the case of random Rayleigh channels as follows.
	
	\noindent \textbf{Proposition 4.} \textit{For $\nu_m^\star>0$ $\forall m \in \mathcal{M}_t$ and $\bh_{mu}\sim\mathcal{CN}(0, \bI)$, the maximum number of sensing streams is limited by}
	\begin{equation}
		\rank(\bF^\star_\mathcal{Q})\leq \max\{M_t-U, 0\}.
	\end{equation}
	\textit{with probability $1$.}
	
	\noindent The proof follows the fact that the probability of any $\bh_u$ drawn from random Gaussian distribution can be spanned by the space constructed by $\bA-\sum_m\nu_m^\star\bD_m$ with probability $0$. Hence, each UE reduces the available dimensions for the sensing stream with probability $1$. 
	
	As shown in Proposition 3, the number of sensing streams upper bounded by the number of APs. More interestingly, in the case of Rayleigh channels, the number of sensing streams is limited to $M_t-U$, i.e., the difference between the number of transmitting APs and UEs. From this result, if the number of transmitting APs is smaller than the number of UEs, no sensing streams is required. However, in the cell-free massive MIMO regime where the number of APs is much larger than the number of UEs, almost a single stream for each AP may be required.
	
	\section{Joint Sensing and Communication: Power Allocation with Fixed Beams} \label{sec:power-allocation}
	In this section, we develop a power allocation formulation. First, we note that the formulated beamforming optimization problem jointly optimizes the power along with the beams since the power constraints of the beams are set to satisfy the power constraints at each AP. Another interesting case with the cell-free massive MIMO, however, is to allocate the power for pre-determined suboptimal beams. For this purpose, in this section, we develop a power allocation formulation for given beams. Differently from the approach developed for power optimization in \cite{behdad2022power}, which adopts an iterative convex-concave programming approach and does not guarantee optimality, we maintain the SDP framework of our paper and develop a power allocation approach with the SDR relaxation. Although this approach cannot provide an optimality guarantee with the relaxation, it can provide an upper-bound with the relaxation.
	
	Mathematically, let us denote the pre-determined unit-power beamforming vectors and power coefficients by $\{\bar{\bff}_{mq}\}$ and $\{p_{mq}\}$. With this notation, the beamforming vectors in the previous formulation in \eqref{eq:JSCSDP} can be written as $\bff_{mq} = \sqrt{p_{mq}} \bar{\bff}_{mq}$. In this model, the fixed beamforming vectors can be selected by the approaches given in \sref{sec:comm-prioritized} and \sref{sec:sensing-prioritized}. With this definition, we can rewrite our JSC objective in terms of the power variables of the beams, $p_{mq}$. Before moving on, we also define the effective channel of UE $u$ and AP $m$ due to the stream $u'$ as $\rho_{muu'} = \bh^H_{mu} \bar{\bff}_{mu'}$ and the sensing channel gain due to the stream $s$ of AP $m$ as $\varrho_{ms} = \absq{\ba^H(\theta_{m}) \bar{\bff}_{m s}} \sum_{m_r \in \mathcal{M}_r} \zeta_{mm_r}^2$. With this, we can define the power allocation problem as
	\begin{subequations} \label{eq:power-allocation}
		\begin{align}
			\textrm{(P3.1a): } \max_{\{p_{ms}\}} \quad & \sum_{m\in \mathcal{M}_t}  \sum_{s \in \mathcal{S}} p_{ms} \varrho_{ms}
			\\
			\begin{split}
				\mathrm{s.t.}\quad & {\gamma_u}^{-1}\absq{\sum_{m \in \mathcal{M}_t} \sqrt{p_{mu}} \rho_{muu}} \\
				&\geq \sum_{s \in \mathcal{S} \backslash \{u\}} \absq{\sum_{m \in \mathcal{M}_t} \sqrt{p_{ms}} \rho_{mus}} + \sigma_{u}^2
			\end{split}
			\\
			&\sum_{s\in\mathcal{S}} p_{ms} \leq P_m, \quad \forall m \in \mathcal{M}_t,
		\end{align}
	\end{subequations}
	This problem, however, is difficult to solve since (i) includes the square root of the power terms, i.e., $\{\sqrt{p_{ms}}\}$, and (ii) contains the summation inside the absolute terms. For (i), we can write the problem in terms of the square root power terms, $\{\sqrt{p_{ms}}\}$. For (ii), we define the per-stream vector form of the power coefficients by stacking the power coefficients of every AP for a given stream, similar to the one applied in \eqref{eq:SINR}, given as  $\bp_s = [\sqrt{p_{1s}}, \ldots, \sqrt{p_{Ms}}]^T$. To complement this variable in our new formulation, we also define the vectors $\bm{\rho}_{us} = [\rho_{1us}, \ldots, \rho_{Mus}]$, and $\bm{\varrho}_{s}=[\varrho_{1s}, \ldots, \varrho_{Ms}]$. In addition, we define the AP selection matrix for the power allocation formulation, $\widetilde\bD_m \in \mathbb{R}^{M\times M}$, where each element of this matrix is given as $[\widetilde\bD_m]_{ij} = 1$ if $i=j=m$, and $0$ otherwise.
	This variable allows rewriting the power constraint in terms of the stacked variable. Then, we can re-write the problem in terms of the newly defined variables as
	\begin{subequations} \label{eq:power-allocation-new_vars}
		\begin{align}
			\textrm{(P3.1b): } \nonumber \\
			\max_{\{\bp_{s}\}} \quad & \sum_{s \in \mathcal{S}} (\bp_{s} \odot \bp_{s})^T \bm{\varrho}_{s}
			\\
			\mathrm{s.t.}\quad & \gamma_u^{-1}\absq{\bp_u^T \bm{\rho}_{uu}} \geq \sum_{s \in \mathcal{S} \backslash \{u\}} \absq{\bp_s^T \bm{\rho}_{us}} + \sigma_{u}^2, \quad \forall u \in \mathcal{U}, 
			\\
			&\sum_{s\in\mathcal{S}} \normsq{\widetilde\bD_m \bp_s} \leq P_m, \quad \forall m \in \mathcal{M}_t, 
		\end{align}
	\end{subequations}
	where the problem is a non-convex QCQP due to the maximization of the sensing SNR, which is a quadratic function of the variable $\bp$. Similar to the \sref{sec:beamforming}, we can transform it into an SDP and apply SDR. For this purpose, we define the new optimization variables for the SDP, i.e., $\bP_s = \bp_s \bp_s^T$, which, by definition, introduces two constraints on the problem (i) The convex symmetric positive semi-definiteness constraint $\bP_s \in \mathbb{S}^+$, where $ \mathbb{S}^+$ is the set of symmetric positive semidefinite matrices, and (ii) the non-convex rank-1 constraint $\rank(\bP_s)=1$. To complement this variable in our formulation, we also define $\bm{\Gamma}_{us}=\bm{\rho}_{us}\bm{\rho}_{us}^H$, and $\bm{\Gamma}'_s = \mathrm{diag}(\bm{\varrho}_{s})$. Then, \eqref{eq:power-allocation-new_vars} can be written as an SDP in terms of these variables, given by
	\begin{subequations} \label{eq:power-allocation-SDP}
		\begin{align}
			\textrm{(P3.1-SDP):} \nonumber\\
			\max_{\{\bP_{s}\}} \quad & \sum_{s \in \mathcal{S}} \Tr{\bP_{s} \bm{\Gamma}'_s} \label{eq:power-allocation-SDP-1}
			\\
			\mathrm{s.t.}\quad & \gamma_u^{-1} \Tr{\bP_u \bm{\Gamma}_{uu}} \geq \sum_{s \in \mathcal{S} \backslash \{u\}} \Tr{\bP_s \bm{\Gamma}_{us}} + \sigma_{u}^2, 
			\\
			&\sum_{s\in\mathcal{S}} \Tr{\bP_s \widetilde\bD_m} \leq P_m, \quad \forall m \in \mathcal{M}_t,
			\\
			&\bP_s \in \mathbb{S}^+ \label{eq:power-allocation-SDP-4}
			\\
			& \rank(\bP_s) = 1
		\end{align}
	\end{subequations}
	The given problem is non-convex due to the rank-1 constraint. To obtain a convex problem, we apply SDR by removing this constraint. Then, the relaxed formulation for the power allocation, denoted as \textbf{(P3.1-SDR)}, can be given by the equations \eqref{eq:power-allocation-SDP-1}-\eqref{eq:power-allocation-SDP-4}. The solution to (P3.1-SDR) can be obtained by the convex solvers. This, however, only results in the matrices $\{\bP^\star_{s}\}$, which is not necessarily rank-1, and the reconstruction of the individual power variables, \{$\bp^\star_s$\}, is required. To reconstruct the solution, one can apply a heuristic approach inspired by the solution in \sref{sec:beamforming} or develop different approaches with the randomization techniques \cite{luo2010semidefinite}. For our purposes of evaluating the beamforming against the power optimization, we utilize a rank-1 heuristic approach, where we take the most significant eigenvector as the solution, and also the solution obtained by the (P3.1-SDR), which provides an upper bound on the power optimization. With the solution completed, we next evaluate our results.
	
	\section{Results}\label{sec:results}

	\begin{figure}[t]
		\centering
		\subfigure[Line setup]{
			\includegraphics[width=.7\columnwidth]{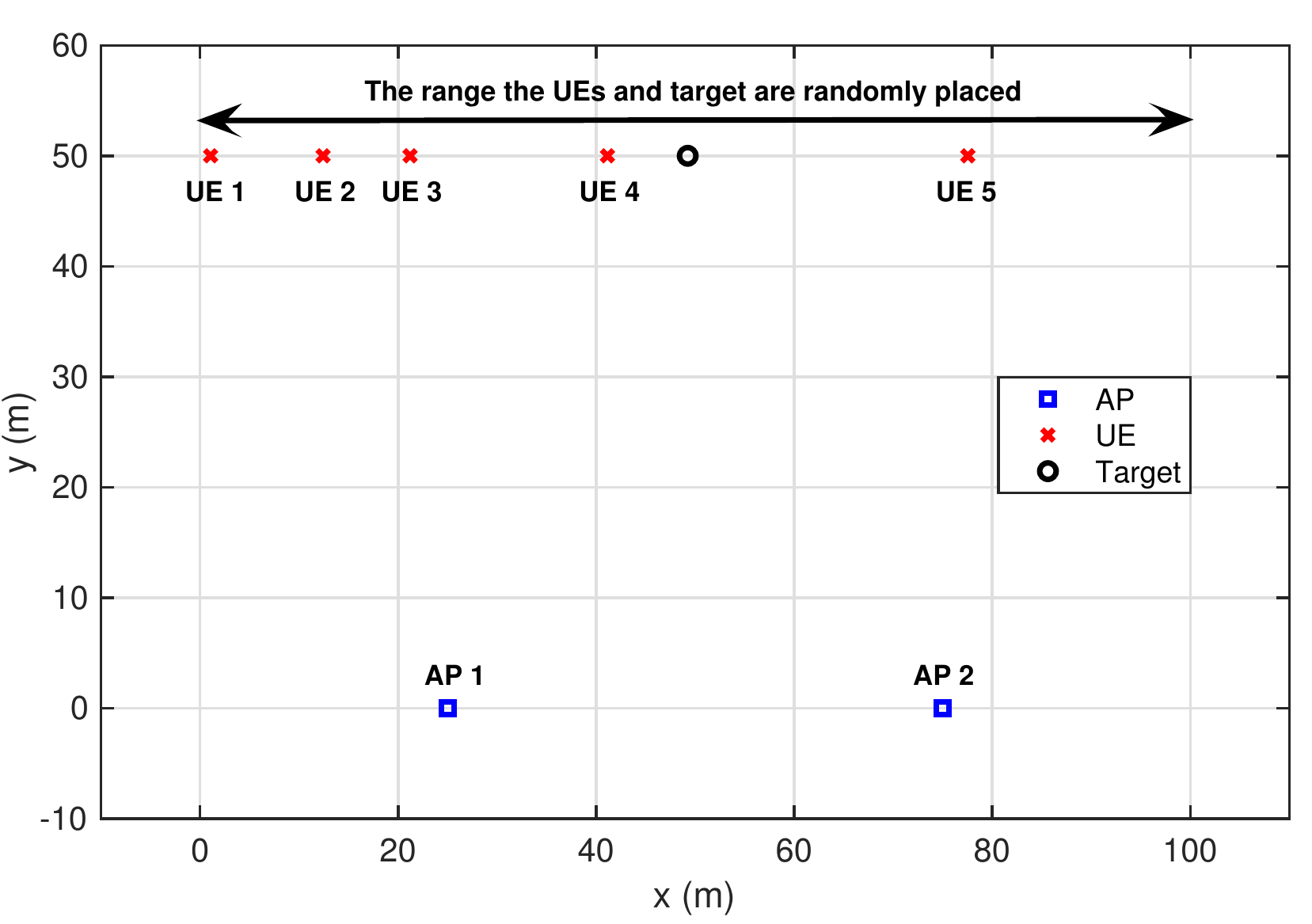}
			\label{fig:simsetup}
		}
		\subfigure[Square setup]{
			\includegraphics[width=.82\columnwidth]{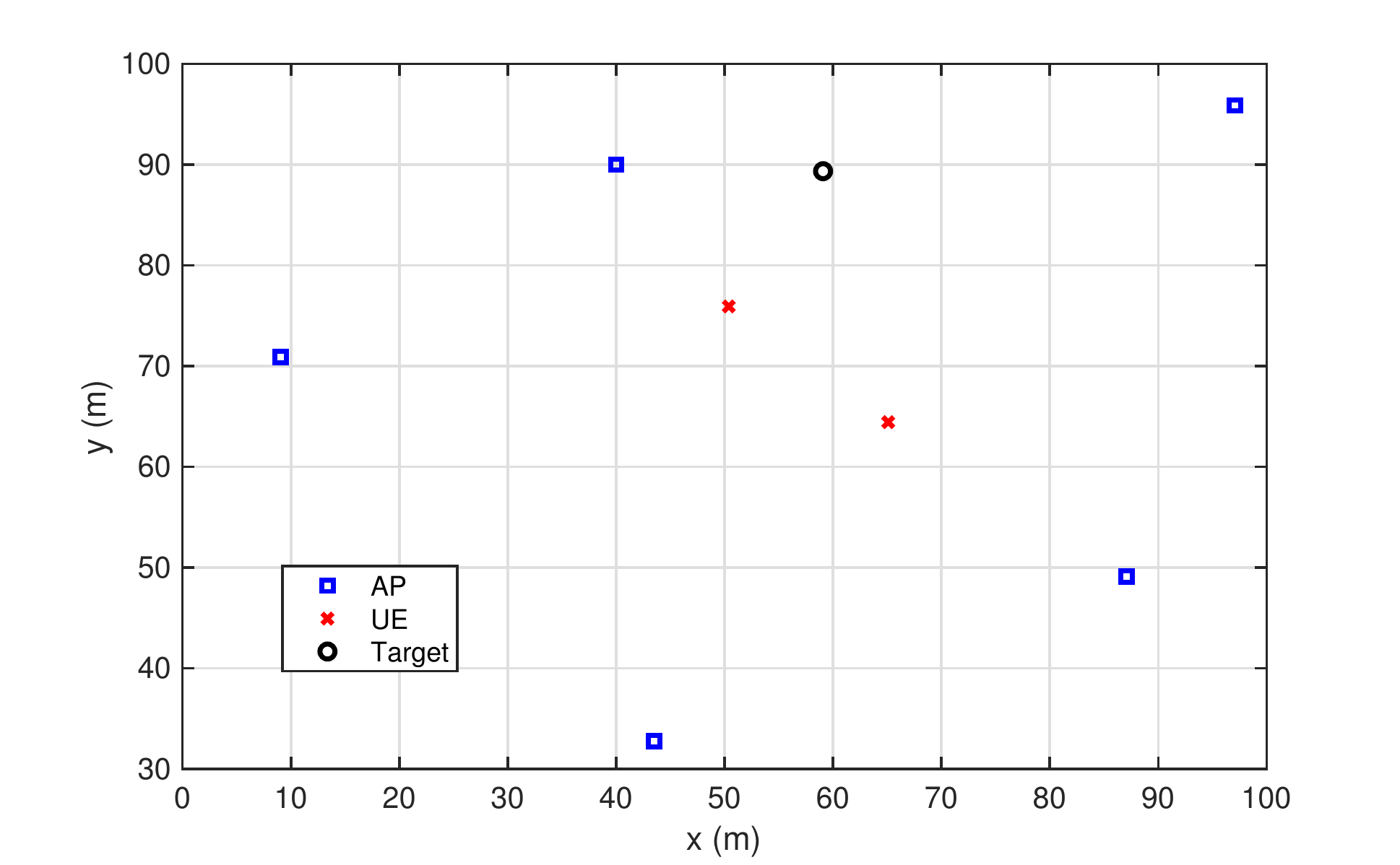}
			\label{fig:simsetup-square}
		}
		\caption{The simulation placement is illustrated. For different realizations, the AP positions are fixed. In (a), the UEs and target are randomly placed over the $y$-axis, while in (b), the UEs and target are randomly placed over the square area of $100$m$\times100$m.} \label{fig:simsetup-all}
	\end{figure}

	In this section, we evaluate the performance of the proposed beamforming solutions for cell-free ISAC MIMO systems \footnote{The implementation is available at ``\url{https://github.com/umut-demirhan/Cell-free-ISAC-beamforming}".}. 
	For this setup, we compare the following solutions: 
	\begin{enumerate} [(i)]
		\item \textbf{NS Sensing - RZF Comm} which designs the sensing beam as conjugate beamforming projected on the null space of the communication channels as in \eqref{eq:sensing-NS}  and implements the communications beams according to the RZF design in \eqref{eq:comm-RZF}. 
		\item \textbf{NS Sensing - OPT Comm} which has the same sensing beam design as in (i) but designs the communication beam based on the max-min optimization in \eqref{eq:SOCP}.
		\item \textbf{CB Sensing - OPT Comm} which first designs the sensing beam as the conjugate beamforming in \eqref{eq:sensing-CB} and then designs the communication beams to solve the max-min optimization in \eqref{eq:SOCP}. 
		\item \textbf{JSC Beam Optimization} which implements the communication and sensing beams based on the SDR problem (P2.1-SDR) and Proposition 1, which jointly optimizes the beamforming vectors based on the communication and sensing functions. 
		\item \textbf{JSC Power Optimization} which implements the communication and sensing beam powers based on the SDR problem in (P3.1-SDR) that jointly optimizes the power coefficients for the given beams along with a rank-1 heuristic approach by taking the most significant eigenvectors as the solution. The pre-determined beamforming vectors are taken as in the (i) \textit{NS Sensing - RZF Comm} approach.
		\item \textbf{JSC Beam SDR UB} which applies the matrix solution for the communication and sensing beams based on the SDR problem in (P2.1-SDR). There is no rank constraint on the beams; hence, it is an upper bound.
		\item \textbf{JSC Power SDR UB} which applies the matrix solution for the communication and sensing beam powers based on the SDR problem (P3.1-SDR). There is no rank constraint on the power variables; hence, it is an upper bound.
	\end{enumerate}
	
	\subsection{LoS Channels}
	In particular, we consider a scenario where $\mathcal{M}_t=\mathcal{M}_r$ with two APs placed at $(25, 0)$ and $(75, 0)$ in the Cartesian coordinates, as shown in \figref{fig:simsetup}. Each AP is equipped with a uniform linear array (ULA) along the $x$ axis of $N_t=N_r=16$ antennas. At $y=50m$, we randomly place one sensing target and the $U=5$ communications users along the $x$-axis. Specifically, the $x$ coordinates of these locations are drawn from a uniform distribution in $[0, 100]$. For the communication channels, we adopt a LOS channel model and take $\sigma^2_u=1$. For the sensing channels, we adopt the parameters $\varsigma^2_{m_r}=1$ and $\zeta_{m_t m_r}=0.1$. The transmit power of the APs is $P_m=0$dBW and the number of sensing streams $Q=1$. In the following, we average the results over $1000$ realizations.
	
	\subsubsection{Providing NS Sensing - OPT Comm SINR for All UEs} 
	With the defined setup, we first focus on an equal rate case, i.e., $\gamma_u$ is the same for every UE. For the selection of this value, we adopt the minimum UE SINR obtained from solution (ii). Further, we do not include the power optimization solution as it is not able to satisfy the SINR constraints for most cases with $\gamma>0.2$.
	
\begin{figure}[!t]
		\centering
		\includegraphics[width=1\linewidth]{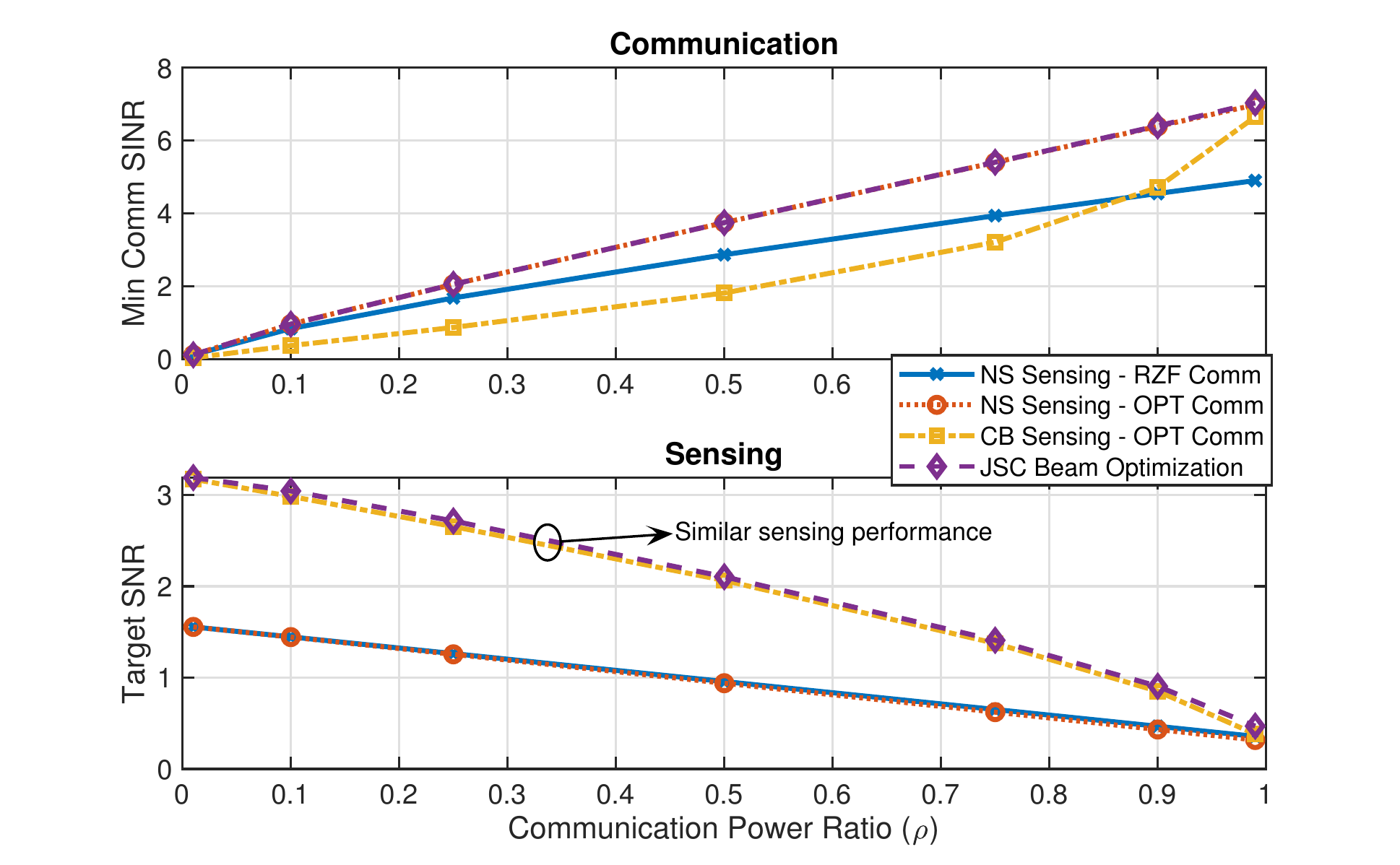}
		\caption{Performance of the solutions for different power allocation ratios for the communications and sensing. The proposed JSC optimization provides a significant gain for sensing while satisfying the best communication SINR.}
		\label{fig:powerscaling}
	\end{figure}

	\textbf{Sensing and Communication Power Allocation:} We first investigate the sensing and communication performance for different power allocation ratios. Specifically, in \figref{fig:powerscaling}, we show the sensing SNR and minimum communication SINR of UEs achieved by the different beamforming solutions for different values of $\rho \in (0, 1)$. It is important to note here that for the beamforming solutions (i)-(iii), the communication and sensing beams are separately designed, and we directly allocate the communication and sensing powers based on the ratio $\rho$. The JSC beam optimization solution (iv) implements the beamforming design in Proposition 1, which optimizes both the structure of the beams and the power allocation. Therefore, and for the sake of comparing with the other approaches, we plot the JSC optimization curve in \figref{fig:powerscaling} by setting the communication SINR threshold to be equal to the achieved SINR by solution (ii). This still respects the total power constraint, which is taken care of by \eqref{eq:powerconstraintJSCSDP}.  
	As seen in the figure, the first two solutions, (i) and (ii), achieve better communication SINR and less sensing SNR compared to the solution (iii). This is expected as the solution (iii) aims to maximize the sensing performance, irrespective of the communication, and hence, it causes some interference to the communication users. 
	\textit{Interestingly, while achieving the best communication performance of the separate solutions, the joint solution provides very similar sensing performance to the MF sensing.} This highlights the gain of the developed JSC beamforming design. 
	
\begin{figure}[!t]
		\centering
		\includegraphics[width=1\linewidth]{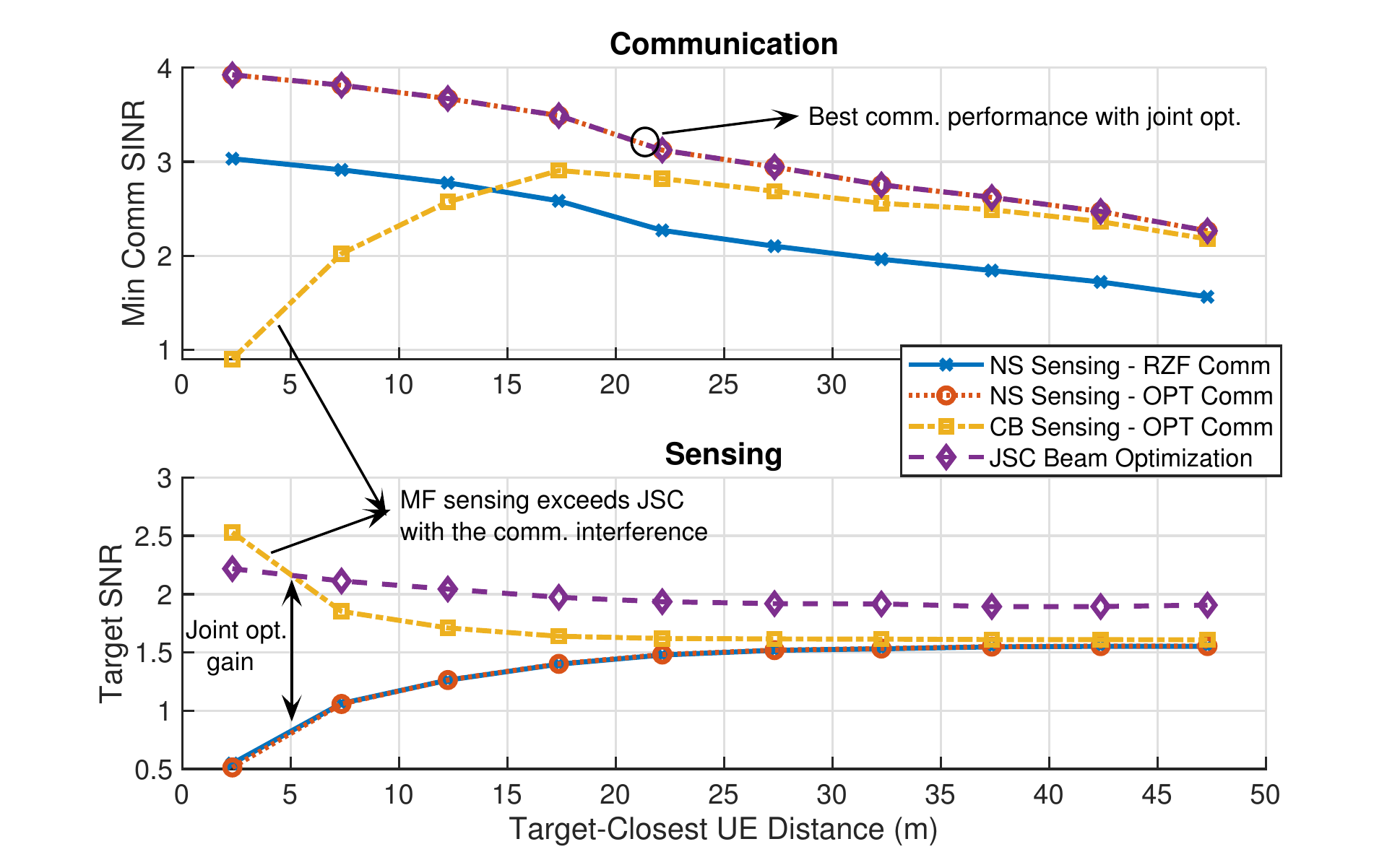}
		\caption{Performance of the solutions versus the distance between the target and closest AP. The proposed JSC optimization provides almost a constant sensing SNR for different distances, with a significant gain over the NS solutions.}
		\label{fig:jscgain}
	\end{figure}

	\textbf{Target distance to closest UE:} To further investigate how the different beamforming approaches impact the trade-off between the sensing and communication performance, we evaluate this performance versus the distance between the sensing target and closest communication UE in \figref{fig:jscgain}. Note that, intuitively, as the sensing target gets closer to the communication users, the overlap between the communication and sensing channels' subspaces increases, which can benefit or penalize the communication and sensing performance depending on the beamforming design. In  \figref{fig:jscgain}, we set the power ratio as $0.5$ for the communication and sensing operation. 
	This figure shows that for the smaller distances/separation between the sensing target and communication users, the conjugate beamforming sensing solution (solution (iii)) optimizes the sensing performance but causes non-negligible interference to the communication, which significantly degrades its performance. 
	On the other side, solutions (i) and (ii), which prioritize the communication and keep the sensing beamforming in the null-space of the communication channels, optimize the communication SINR and degrade the sensing SNR. For the SINR constraint of the JSC optimization, we again adopt the SINR obtained by solution (ii), which achieves the best communication performance. Hence, the achieved communication SINR of this solution and JSC  beam optimization are the same. \textit{The sensing SNR, however, enjoys the advantage of the joint beam optimization. Specifically, it provides almost a constant sensing performance for different target-closest UE distances: Achieving a close sensing performance to solution (i) when the separation between the sensing target and communication users is small and exceeds the performance of all the other three solutions when this separation is large, which highlights the potential of the joint beamforming design.}

	\subsubsection{Providing NS Sensing - RZF Comm SINR for each UE} 
\begin{figure}[!t]
		\centering
		\includegraphics[width=1\linewidth]{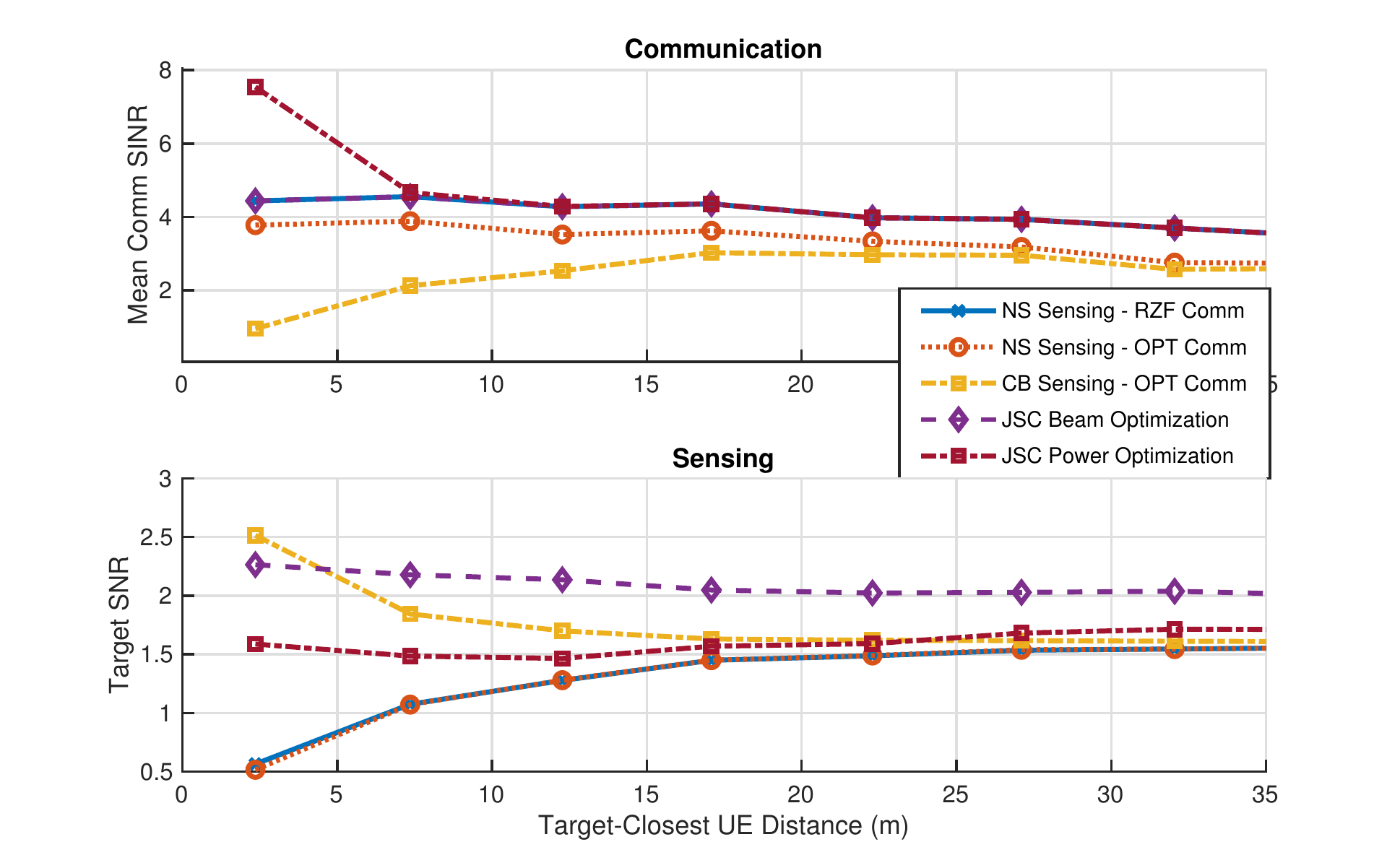}
		\caption{Performance of the mean SINR versus the distance between the target and closest AP. The optimizations are carried out to satisfy individual rates achieved by RZF. A similar pattern to the previous case is observed.}
		\label{fig:jscmeangain}
	\end{figure}

Now, to further investigate the performance of the joint optimization, we select each $\gamma_u$ individually as their SINR obtained from solution (i). Differently from the previous approach (minimum RZF rate), we test the imbalanced rates and the gain concerning solution (i) while satisfying the same SINR values. In \figref{fig:jscmeangain}, we provide the closest distance figure with these SINR constraints. In the figure, the beamforming (with $Q=0$) and power solutions with corresponding upper bounds, (iv) with (vi) and (v) with (vii), achieve the same results, hence only (iv) and (v) are illustrated. Compared to the previous figure, the mean SINR of solution (i) shows less degradation with larger distances, thanks to being able to exploit the imbalance for the mean rates. On the other hand, the beamforming optimization (iv) achieves the same SINR and provides a significant sensing SNR gain over all other solutions without any sensing beams, as indicated by Proposition 4. This shows a similar advantage to the previous case. For very small target-closest UE distances, (iv) the beamforming optimization is worse for sensing than (iii) CB Sensing - OPT Comm, which cannot achieve similar communication SINRs due to the high interference. At the communication part, (v) the power optimization solution achieves higher average communication SINR with very close distances since allocating the power onto the UE with the closest distance provides more gain for sensing than the NS sensing beam. In the general case, however, (iv) the beamforming optimization provides significant sensing gain over all the solutions, showing a similar pattern to the previous case. 
	
	\subsection{Rayleigh Channels}
	As we have only investigated a simplified setup so far to examine the effects, we now provide a more realistic setup. In this setup, the AP and UEs are placed over a square area of $100$m$\times100$m. We utilize the $f_c=28$GHz band and place $M_t=M_r=5$ APs, each equipped with a ULA of $N_t=8$ antennas. To show the need for additional sensing streams, based on our observations in \sref{sec:beamforming}, we take the number of UEs as $2$. The setup is illustrated in \figref{fig:simsetup-square}. For the path-loss, we adopt the 3GPP UMi path loss \cite{3GPP2017} given as $\textrm{PL}=-32.4 - 21\log_{10}(\textrm{distance}) - 20\log_{10}(f_c)$. Further, we assume Rayleigh fading for the AP-UE channels. The receiver noise at the UEs is taken as $-135$dBm.
	
\begin{figure}[!t]
		\centering
		\includegraphics[width=1\linewidth]{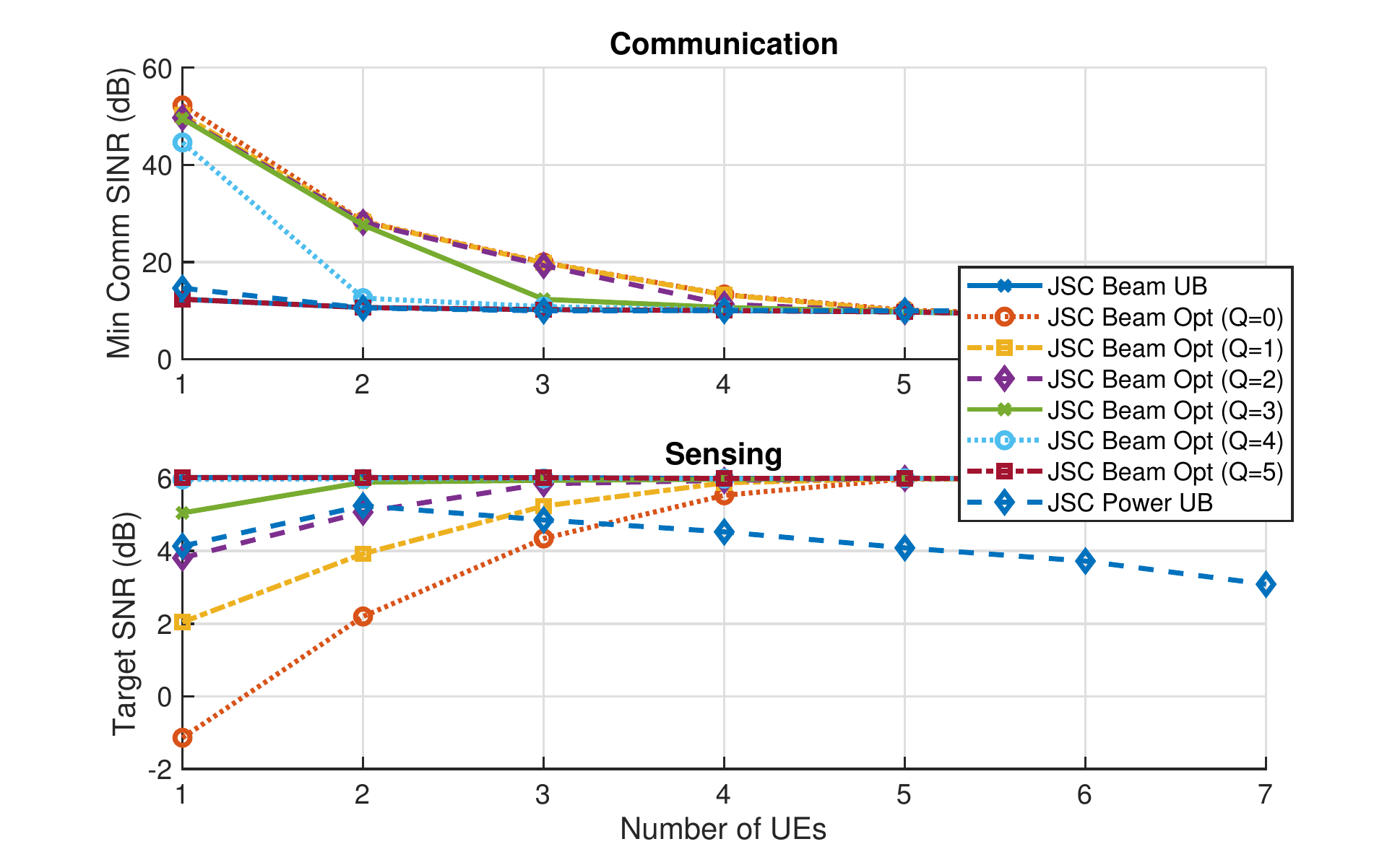}
		\caption{Performance of the SDR-based optimization solutions with varying number of sensing streams and number of UEs.}
		\label{fig:jscnumue}
	\end{figure}

	To evaluate the performance in this setup, we set the minimum communication SINR threshold as $10$dB for all UEs.
	Within the evaluations, we only use solutions beam and power optimization solutions, as they are able to conform to the SINR constraints while maximizing the sensing SNR. In \figref{fig:jscnumue}, we show the achieved sensing SNR and communication SINR values with different numbers of UEs. As expected from Proposition 4, we need $Q=M-U$ streams to achieve the beamforming upperbound. To that end, the sensing SNR provided by any beamforming optimization (iv) curve with $Q\geq M-U$ achieves the same value, while satisfying the communication constraints. The power optimization can exceed the solution obtained by a single sensing beam with $U\leq 2$, which shows the case it may be preferable if the sensing streams are limited, and there are more APs than UEs. This result also highlights the advantage of the beamforming solution. In addition, we note that the provided heuristic rank-1 optimization does not provide sufficient SNR, and better heuristic methods for the rank-1 constructions are required. We, however, leave this as a future work for researchers as our focus of the paper is to investigate JSC beamforming, and also compare it with the potential of power allocation. As shown in the figure, adding sensing streams can provide advantages for the cell-free massive MIMO systems, and this can allow further gain over the suboptimal beams with power allocation.
	
	\section{Conclusion}\label{sec:conclusion}
	In this paper, we investigated downlink beamforming for joint sensing and communication in cell-free massive MIMO systems. Specifically, we designed communication-prioritized sensing beamforming and sensing-prioritized communication beamforming solutions as the baseline. Further, we have developed an optimal solution for the JSC beamforming. The results showed the advantage of the joint optimization, where the developed JSC beamforming is capable of achieving nearly the SINR of the communication-prioritized sensing beamforming solutions with almost the same sensing SNR of the sensing-prioritized communication beamforming approaches.

	\appendix
	
	\subsection{Derivation of the Sensing SNR} \label{appendix-sensingSNR}
	The expectation of the nominator can be simplified as
	\begin{align}
		&\bbE\left[ \sum_{m_r \in \mathcal{M}_r} \frobeniusq{ \overline\bG_{m_r} \overline\bX}\right]
		\\
		&=\sum_{m_r \in \mathcal{M}_r}  \textrm{Tr}\left\{\bbE\left[ \overline\bX \, \overline\bX^H \overline\bG^H_{m_r} \overline\bG_{m_r} \right\}\right]  \label{eq:appendixa-eq1}
		\\
		&=\sum_{m_r \in \mathcal{M}_r}  \textrm{Tr}\bigg\{\underbrace{\bbE\left[ \overline\bX \, \overline\bX^H \right]}_{=L \cdot\bI}  \bbE\left[ \overline\bG^H_{m_r} \overline\bG_{m_r} \right] \bigg\}  \label{eq:appendixa-eq2} \\
		\begin{split}  \label{eq:appendixa-snrl}
			&=L \sum_{m_r \in \mathcal{M}_r} \sum_{m_t\in \mathcal{M}_t} \bigg( \textrm{Tr} \, \bbE\left[ \overline\bG_{m_t m_r} \overline\bG^H_{m_t m_r}
			\right] \\
			&\quad + L \sum_{m'_t \in \mathcal{M}_t \backslash \{m_t\}} \textrm{Tr} \, \bbE\left[ \overline\bG_{m_t m_r} \overline\bG^H_{m'_t m_r}
			\right] \bigg) 
		\end{split}
	\end{align}
	where \eqref{eq:appendixa-eq1} and  \eqref{eq:appendixa-eq2} are obtained by applying the expansion of the Frobenius norm, interchanging expectation and trace, and permutating the inner terms of the trace operation several times. To obtain \eqref{eq:appendixa-snrl}, we apply the definition $\overline{\bG}_{m_r}=\sum_{m_t} \overline{\bG}_{m_t m_r}$ given in \eqref{eq:simplified_form}, and re-organize the multiplication terms. Further, for \eqref{eq:appendixa-snrl}, due to the expectation over the random variables $\{\alpha_{m_t m_r}\}$ and independence of them, we have $\bbE\left[ \overline\bG_{m_t m_r} \overline\bG^H_{m'_t m_r} \right] = 0$, which makes the latter line of \eqref{eq:appendixa-snrl} zero. For the former, we have
	\begin{align}
		&\textrm{Tr} \, \bbE\left[ \overline\bG_{m_t m_r} \overline\bG^H_{m_t m_r}
		\right] \\
		& = \textrm{Tr} \, \bbE\left[ \absq{\alpha_{m_t m_r}} \ba(\theta_{m_r}) \ba^H(\theta_{m_t}) \overline{\bF}_{m_t} \overline{\bF}^H_{m_t} \ba(\theta_{m_t}) \ba^H(\theta_{m_r})   \right] \\
		&= \zeta_{m_t m_r} ^2 \textrm{Tr} \, \bbE\left[ \ba^H(\theta_{m_t}) \overline{\bF}_{m_t} \overline{\bF}^H_{m_t} \ba(\theta_{m_t}) \underbrace{\ba^H(\theta_{m_r}) \ba(\theta_{m_r}) }_{=N_r}  \right] \\
		&= \zeta_{m_t m_r}^2 N_r \normsq{\ba^H(\theta_{m_t}) \overline{\bF}_{m_t}}. \label{eq:appendix-sensingSNR-nom2}
	\end{align}
	For the denominator, we can write
	\begin{equation} \label{eq:appendix-sensingSNR-denom}
		\begin{split}
			\bbE\left[ \sum_{m_r \in \mathcal{M}_r} \frobeniusq{\bN_{m_r}}\right] &= \bbE\left[ \sum_{m_r \in \mathcal{M}_r} \sum_{\ell=1}^L \normsq{\bn_{m_r}[\ell]}\right] 
			\\
			&= L N_r \sum_{m_r \in \mathcal{M}_r} \varsigma_{m_r}^2. 
		\end{split}
	\end{equation}
	Finally, combining \eqref{eq:appendixa-snrl}, \eqref{eq:appendix-sensingSNR-nom2}, and \eqref{eq:appendix-sensingSNR-denom} in \eqref{eq:SNRsensing}, we obtain the result.
	
	\subsection{Derivation of the Sensing SNR in SDP form} \label{appendix-sensingSNR-SDP}
	
	To simplify the sensing SNR expression given in \eqref{eq:SNRsensing} in the SDP form, we can write the nominator as
	\begin{align}
		&\bbE\left[ \sum_{m_r \in \mathcal{M}_r} \normsq{ \overline\bG_{m_r} \overline\bX}\right] 
		\\
		&=  N_r L \sum_{m_t\in \mathcal{M}_t} \sum_{m_r \in \mathcal{M}_r} \zeta_{m_t, m_r}^2 \sum_{s \in \mathcal{S}} \absq{\ba^H(\theta_{m_t}) \bff_{m_t s}}
		\\
		&=  N_r L \sum_{m_t\in \mathcal{M}_t} \sum_{m_r \in \mathcal{M}_r} \sum_{s  \in \mathcal{S}} \absq{\zeta_{m_t, m_r} \overline\ba^H \bD_{m_t} \bff_{s}} \label{eq:SNRsens1}
		\\
		&=N_r L \sum_{m_t\in \mathcal{M}_t} \sum_{m_r \in \mathcal{M}_r} \sum_{s  \in \mathcal{S}}\Tr{\zeta_{m_t, m_r}^2 \overline\ba^H \bD_{m_t} \bff_{s} \bff_{s}^H \bD_{m_t} \overline\ba} \label{eq:SNRsens2}
		\\
		&= N_r L \sum_{m_t\in \mathcal{M}_t} \sum_{m_r \in \mathcal{M}_r} \sum_{s  \in \mathcal{S}}\Tr{\zeta_{m_t, m_r}^2 \bD_{m_t} \overline\bA \bD_{m_t} \bF_{s} } \label{eq:SNRsens3}
		\\
		&= N_r L \Tr{ \sum_{m_t\in \mathcal{M}_t} \bD_{m_t} \overline\bA \bD_{m_t} \bar{\zeta}_{m_t} \sum_{s  \in \mathcal{S}} \bF_{s} } \label{eq:SNRsens5}
		\\
		&= N_r L  \Tr{ \bA \sum_{s  \in \mathcal{S}} \bF_{s} } \label{eq:SNRsens6}
	\end{align}
	where we obtain \eqref{eq:SNRsens1} by the definitions of $\overline\ba$, $\bD_{m_t}$, and $\bff_s$, \eqref{eq:SNRsens2} by $|\bX|^2 = \Tr{\bX\bX^H}$, \eqref{eq:SNRsens3} by cyclic permutation property of the trace operation and the definitions of $\bF_s$ and $\overline\bA$, \eqref{eq:SNRsens5} by rearranging the summations and defining $\bar{\zeta}_{m_t}=\sum_{m_r \in \mathcal{M}_r} \zeta_{m_t, m_r}^2$, and \eqref{eq:SNRsens6} by defining $\bA=\sum_{m_t\in \mathcal{M}_t} \bar{\zeta}_{m_t} \bD_{m_t} \overline\bA \bD_{m_t}$. 
	
	\subsection{Proof of Proposition 1} \label{sec:appendixb}
	This extends the proof in \cite{liu2020joint}. For this purpose, we first note that in the problem formulation in (P2.1-SDR), the sensing variable, $\bF_\mathcal{Q}$, are utilized together as a summation of all of the streams, both in the objective and constraints. Hence, if we define $\bar{\bF} = \sum_{s\in\mathcal{S}} \bF_s$, we can eliminate the sensing term $\bF_\mathcal{Q}$, and apply the optimization in terms of the user streams $\bF_u$ and $\bar{\bF}$. To that end, we re-formulate the problem (P2.1-SDR) as
	\begin{subequations}
		\begin{align}
			\max_{\{\bF_{u}\}, \bar{\bF}} \quad & { \Tr{ \bA \bar{\bF} } }
			\\ &
			\left(1+\gamma_u^{-1}\right) \Tr{\bQ_{u} \bF_u } - \Tr{\bQ_{u} \bar{\bF}} \geq \sigma_{u}^2, \quad \forall u \in \mathcal{U} \label{eq:appendix-SINR}
			\\ &
			\Tr{\bD_m \bar{\bF}} = P_m, \quad \forall m \in \mathcal{M}_t,  \label{eq:appendix-power}
			\\ & 
			\bF_u \in \mathbb{S}^+, \quad \forall u \in \mathcal{U}, \label{eq:appendix-usersemidefinite}
			\\ & 
			\bar{\bF} - \sum_{u \in \mathcal{U}} \bF_u \in \mathbb{S}^+, \quad 	\bar{\bF} \in \mathbb{S}^+. \label{eq:appendix-sum}
		\end{align}
	\end{subequations}
	Let us denote the variables obtained by the solution of this problem by $\{\bF'_u\}$ and $\bar{\bF}'$. Using this solution, we aim to construct an alternative optimal solution of rank-1. For this purpose, we construct the following rank-1 set of solutions
	\begin{equation}\label{eq:proof-defs}
		\bar{\bF}''=\bar{\bF}', \quad \bF''_u=\bff''_u(\bff''_u)^H, \quad \bff''_u=(\bh_u^H \bF'_u \bh_u)^{-\frac{1}{2}} \bF'_u \bh_u.
	\end{equation}
	whose optimality needs to be proved. For this, we need to check if (i) the value of the objective is the same and (ii) the constraints are satisfied. First, the objective only contains the summation variable and provides the optimal value by definition. For \eqref{eq:appendix-SINR}, we define $v_u=(\bh_u^H \bF'_u \bh_u)^{-\frac{1}{2}}$, and write $\Tr{\bQ_u \bF_u''} = \Tr{\bh_u v_u^2 \bh_u^H \bF_u' \bh_u \bh_u^H \bF_u'^H} = \Tr{\bQ_u \bF_u'}$,
	where we used the cyclic permutation property of the trace and $\bF_u'^H = \bff_u'\bff_u'^H = \bF_u'$. With the addition of $\bar{\bF}''=\bar{\bF}'$, \eqref{eq:appendix-SINR} is satisfied. Similarly, the constraints \eqref{eq:appendix-power} and \eqref{eq:appendix-sum} are already satisfied by $\bar{\bF}''=\bar{\bF}'$. Further, \eqref{eq:appendix-usersemidefinite} and the solution being rank-1 are also satisfied by the definition of ${\bF''_u}$ in \eqref{eq:proof-defs}. For \eqref{eq:appendix-sum}, we have $\bv^H (\bF'_u-\bF''_u) \bv =  \bv^H \bF'_u \bv -  (\bh_u^H \bF'_u \bh_u)^{-1} \absq{\bv^H \bF'_u \bh_u}$.	From the Cauchy-Schwarz inequality, we also have $(\bv^H \bF'_u \bv)(\bh_u^H \bF'_u\bh_u) \geq \absq{\bv^H \bF'_u \bh_u}$. Combining these two equations, we obtain $\bv^H (\bF'_u-\bF''_u) \bv \geq 0$, which leads to $\bv^H \bF''_u \bv \geq 0$ since it is the summation of two semidefinite matrices, $\bF'_u-\bF''_u$ and $\bF'_u$. Finally, \eqref{eq:appendix-sum} can be shown via $\bar{\bF}'' - \sum_{u\in\mathcal{U}}\bF''_u = \bar{\bF}' - \sum_{u\in\mathcal{U}}\bF'_u + \sum_{u\in\mathcal{U}}(\bF'_u - \bF''_u)$ which again leads to the summation of semidefinite matrices.
	
	Finally, for constructing the sensing matrices of the solution, we want to find $Q$ rank-1 matrices whose summation is $\sum_{q\in\mathcal{Q}}\bF''_q$. For this purpose, we can utilize the eigendecomposition, i.e., $\sum_{Q} \bF''_q = \bU \Lambda \bU^H = \sum^{Q'}_{q'=1} \lambda_{q'} {\bu}_{q'}{\bu}_{q'}^H$, and take the largest $Q$ eigenvectors as the beams via $\bff''_q = \sqrt{\lambda_{u'}} \, {\bu}_{q'}$. Here, it is important to note that it is only possible if the rank of the summation, $Q'=\mathrm{rank}(\sum_{q\in\mathcal{Q}}\bF''_q)$, is smaller than or equal to the number of the sensing streams, $Q$.
	
	\subsection{Derivation of D2.1-SDR}\label{sec:appendixc}
	
	For the dual of (P2.1-SDR), we first write the Lagrangian function as follows.
	\begin{align}
		\begin{split}
			&\mathcal{L}(\{\bF_u\}, \bF_\mathcal{Q},  \{\bZ_u\}, \bZ_\mathcal{Q}, \{\lambda_u\}, \{\nu_m\}) 
			\\
			& = \sum_u \Tr{\bA \bF_u} + \Tr{\bA {\bF}_\mathcal{Q}} + \sum_u \lambda_u \left(1+\gamma_u^{-1}\right) \Tr{\bQ_{u} \bF_u } 
			\\
			&\quad - \sum_{u} \sum_{u'} \lambda_{u'} \Tr{\bQ_{u'} \bF_u} - \sum_{u'} \lambda_{u'} \Tr{\bQ_{u'} {\bF}_\mathcal{Q}} 
			\\
			&\quad - \sum_u \lambda_u \sigma_{u}^2  - \sum_u \sum_m \nu_m \Tr{\bD_m \bF_u} - \sum_m \nu_m \Tr{\bD_m {\bF}_\mathcal{Q}} 
			\\
			&\quad + \sum_m \nu_m P_m  + \sum_u\Tr{\bZ_u \bF_u} + \Tr{\bZ_\mathcal{Q} {\bF}_\mathcal{Q}}
		\end{split}
	\end{align}
	where $\{\lambda_u\}\geq0$, $\{\nu_m\}\geq0$, $\{\bZ_u\} \succeq 0$, and $\bZ_\mathcal{Q} \succeq 0$ are the Lagrangian variables corresponding to the SINR constraints, AP power constraints, and the semidefiniteness constraints for the user matrices and the sensing matrix. Collecting all the multiplications with $\bF_u$, and ${\bF}_\mathcal{Q}$, we can rewrite the Lagrangian function in a compact form as
	\begin{align}
		\begin{split}
			&\mathcal{L}(\{\bF_u\}, \bF_\mathcal{Q}, \{\bZ_u\}, \bZ_\mathcal{Q}, \{\lambda_u\}, \{\nu_m\})  \\
			&= \sum_m \nu_m P_m - \sum_u \lambda_u \sigma_{u}^2 
			\\
			&\quad + \sum_u \Tr{(\bB_u + \bZ_u) \bF_u} + \Tr{(\bB_\mathcal{Q} + \bZ_\mathcal{Q}) {\bF}_\mathcal{Q}}.
		\end{split}
	\end{align}
	Then, we note that supremum of Lagrangian for $\bF_u$ and $\bF_q$ is only bounded if $\bB_u + \bZ_u = \b0$. Thus, replacing the variable $\bZ_u \geq 0$ with $\bB_u \leq 0$, and similarly, for the sensing matrix, we can derive the dual problem via
	\begin{align}
		\begin{split}
			&\min \sup_{\{\bff_u\}, \bff_q} \mathcal{L}(\{\bff_u\}, \bar{\bff}_q, \{\lambda_u\}, \{\nu_m\})  = \\
			&
			\begin{cases}
				\sum_m \nu_m P_m - \sum_u \lambda_u \sigma_u^2 & \text{if }\bB_q \preceq 0 \text{ and } \bB_u \preceq 0,\ \forall u \in \mathcal{U}
				\\
				\infty & \text{otherwise.}
			\end{cases}
		\end{split}
	\end{align}
	
	\balance
	% Generated by IEEEtran.bst, version: 1.14 (2015/08/26)

\end{document}